\documentclass[aps,prl,reprint,superscriptaddress]{revtex4-1}

\usepackage{graphicx}
\usepackage{mathtools}
\usepackage{epstopdf}
\usepackage{amsmath}    
\usepackage{amsfonts}
\usepackage{verbatim}   
\usepackage{color}      
\usepackage{subfigure}  
\usepackage{soul}
\usepackage{xr}

\newcommand\numberthis{\addtocounter{equation}{1}\tag{\theequation}}
\begin{document}


\title{Biochemical machines for the interconversion of mutual information and work}


\author{Thomas McGrath}
\author{Nick S. Jones}
\affiliation{Department of Mathematics, Imperial College London, London, SW7 2AZ, UK}
\author{Pieter Rein ten Wolde}
\affiliation{FOM
  Institute AMOLF, Science Park 104, 1098 XE Amsterdam, The
  Netherlands}
\author{Thomas E. Ouldridge}
\email{t.ouldridge@imperial.ac.uk}
\affiliation{Department of Bioengineering, Imperial College London, London, SW7 2AZ, UK}


\date{\today}

\begin{abstract}
We propose a physically-realizable information-driven device consisting of an enzyme in a chemical bath, interacting with pairs of molecules prepared in correlated states. These correlations persist without direct interaction and thus store free energy equal to the mutual information. The enzyme can harness this free energy, and that stored in the individual molecular states, to do chemical work. Alternatively, the enzyme can use the chemical driving to create mutual information. A modified system can function without external intervention, approaching biological systems more closely.
\end{abstract}

\pacs{}

\maketitle


Organisms exploit correlations in their environment to survive and grow. This  fact holds across scales, from bacterial chemotaxis, which leverages the spatial clustering of food molecules \cite{Micali2016,Becker2015}, to the loss of leaves by deciduous trees, which is worthwhile because sunlight exposure is highly correlated from day to day. Evolution itself relies on correlations across time and space, otherwise a mutation which is beneficial would immediately lose its utility and selection would be impossible. 

Biological systems also generate correlations. In particular, information transmission is an exercise in correlating input and output \cite{Shannon1949}, and
recent years have thus seen information theory applied to biological systems involved in sensing \cite{Govern2014,Barato2014,Ouldridge2015}, signalling \cite{Cheong2011,deRonde2011}, chemotaxis \cite{Micali2016,Becker2015}, adaption \cite{Sartori2014, Ito2014} and beyond \cite{Bialek2015}. In the language of information theory, correlated variables $X$ and $Y$ have a positive mutual information $I(X;Y) = \sum_{x\in X, y\in Y}p(x,y)\ln\frac{p(x,y)}{p(x)p(y)}$ (measured in nats), with $p(x,y)$ the joint probability of a given state and $p(x)$, $p(y)$ the marginals. The ``information entropy" $H(Y) = - \sum_{y \in Y} p(y) \ln p(y)$ quantifies $Y$'s uncertainty, and $I(X;Y)$ is the reduction in this entropy given knowledge of $X$: $I(X;Y) = H(Y) - H(Y|X)$. The mutual information is symmetric, non-negative, and zero if and only if  $X$ and $Y$ are statistically independent.

Information theory is also deeply connected to thermodynamics \cite{Landauer1961,Bennett1982, Sagawa2009, Bauer2012, Still2012, Sagawa2012, Horowitz2014, Barato2014b,  Parrondo2015}. Sagawa and Ueda \cite{Sagawa2009}, building on \cite{Landauer1961,Bennett1982}, showed that measurement cycles have a minimal work cost equal to the mutual information generated between data and memory. Horowitz and Esposito showed that entropy production within a system $X$ can be negative if $X$ is coupled to a second system $Y$, and transitions in $X$ decrease $I(X;Y)$ \cite{Horowitz2014}. A third key result, essential to exorcising Maxwell's Demon \cite{Fahn1996,Ouldridge2015}, is that if $X$ and $Y$ are uncoupled from each other, yet coupled to heat baths at temperature $T$, then the total free energy is \cite{Horowitz2013, Esposito2011}
\begin{equation}
\tilde{F}(X,Y) = \tilde{F}(X) + \tilde{F}(Y) + k_BTI(X; Y).
\label{eq:info}
\end{equation}
Here $\tilde{F}(X) = F_{\rm eq}(X) - kT\sum_{x\in X}p(x)\ln(p_{\rm eq}(x)/p(x))$ is the non-equilibrium free energy \cite{Parrondo2015,Esposito2011}, with the tilde indicating the generalisation from the standard equilibrium free energy $F_{\rm eq}(X)$. Systems $X$ and $Y$ could be two non-interacting spins, or two physically separated molecules. For uncoupled systems, the partition function is separable and $X$ and $Y$ are independent in equilibrium, $I_{\rm eq}(X;Y)=0$. However, correlations induced by coupling between $X$ and $Y$ at earlier times could persist even after the coupling has been removed \cite{Ouldridge2015}, or $X$ and $Y$ may have been initialized by related processes, allowing $I(X;Y)>0$. If $I(X;Y)>0$ despite the current absence of interactions, $\tilde{F}(X,Y)>F_{\rm eq}(X,Y) = F_{\rm eq}(X) + F_{\rm eq}(Y)$ even if $\tilde{F}(X) = F_{\rm eq}(X)$ and $\tilde{F}(Y) = F_{\rm eq}(Y)$. Thus $I(X;Y)>0$ between uncoupled systems  implies  excess free energy, and excess free energy is a resource from which work can be extracted \cite{Parrondo2015,Esposito2011}.

Information appears inherently abstract and work-performing devices coupled to strings of 0s and 1s \cite{Bennett1982,Mandal2012,Mandal2013,Barato2013,Lu2014,Cao2015,Boyd2016,Horowitz2013} can seem remarkable. However, as Landauer pointed out, ``information is physical" \cite{Landauer1991}, and processing it calls for a physical realisation. 
We propose biochemical information-exploiting devices to show both thermodynamically (via Eq.\,\ref{eq:info}) and physically (via the actual information-processing mechanism) how mutual information can be used to do work, setting the basis for more sophisticated information-exploiting devices.

Our first device is a ``tape-driven" biochemical  machine, which, unlike previous tape-driven devices \cite{Bennett1982,Mandal2012,Mandal2013,Barato2013,Lu2014,Cao2015,Boyd2016,Horowitz2013}, exploits {\it mutual} information within the input. The information-processing mechanism is explicit, differing from measurement-feedback devices previously considered \cite{Bauer2012, Horowitz2013, toyabe2010experimental, Vidrighin2016}. A recent study did suggest work extraction from a perfectly-correlated quantum tape \cite{Chapman2015}. However, like the majority of tape-driven machines \cite{Bennett1982, Mandal2012,Mandal2013,Barato2013,Cao2015,Boyd2016,Horowitz2013}, the dynamics is not based on an actual physical system. It is also unclear how the initial state would be created, or the mechanism generalised \cite{Chapman2015}. Using an explicit system emphasises the constraints under which information-processing devices operate. For example, the consequences of externally controlling a tape that physically couples to the device are often ignored. With these effects in mind, we introduce a modified device that functions without a control, removing implicitly neglected costs, providing a simpler route to constructing an actual device, and moving closer to autonomous biological systems.

Our `engine' is an enzyme $E$, which is converted from inactive ($E$) to active ($E^\dagger$) by the binding of molecule $Y$. The enzyme can also bind to nucleotides ADP/ATP, and a substrate  $X$. When active, $E^\dagger$ catalyses phosphate  exchange between ATP and $X$ 
\begin{equation}\label{eq:chem1}
X + {\rm ATP} +E^\dagger \rightleftharpoons E^\dagger{\textrm{-}}X{\textrm{-ADP-P}}_{\rm i} \rightleftharpoons E^\dagger +X^{*} + {\rm ADP}.
\end{equation}
One natural  enzyme/substrate/activator combination would be LBK1/AMPK/STRAD-MO25 \cite{Zeqiraj2009}, although others exist \cite{Cowan-Jacob2014} and engineered examples \cite{Qiao2006,Karginov2010} might be  optimal. The active enzymatic engine transfers free energy from substrates (the fuel) to a bath of ATP/ADP (the load), or vice-versa. The work done in converting a single ADP into ATP is given by the difference in chemical potentials, $\mu_{\rm ATP}-\mu_{\rm ADP}$. In dilute solutions,  $\mu_{\rm ATP}-\mu_{\rm ADP} =  \mu^0_{\rm ATP} - \mu^0_{\rm ADP} + k_{\rm B}T \ln([{\rm ATP}]/[{\rm ADP}])$, with $\mu^0_{\rm ATP} - \mu^0_{\rm ADP}$ the ``intrinsic" contribution.  Changing $[{\rm ATP}]/[{\rm ADP}]$ thus adjusts the load.

\begin{figure}
\includegraphics[width=0.45\textwidth]{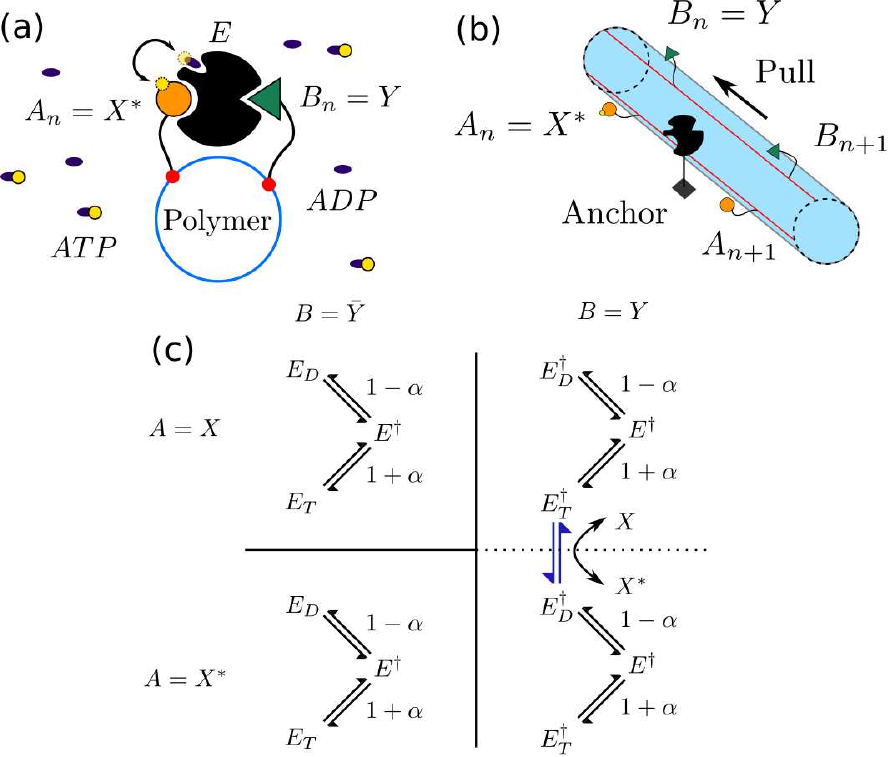}
\caption{(a) Schematic showing an enzyme interacting with the $n^{\rm th}$ pair of sites, catalysing $X^* + ADP \rightleftharpoons X + ATP$ on tape $A$ due to the presence of activator $Y$ on tape $B$. Tapes extend into and out of page, and yellow circles indicate phosphorylation. (b) Two tapes, coordinated on a single polymer, being pulled past a tethered enzyme. We propose a DNA origami-based construction in Ref.\,\cite{SI}, Section 1. (c) State space of the model. $E_T$ and $E_D$ correspond to ATP- and ADP-bound enzymes, respectively; $E_-$ to a free enzyme. All unlabelled transitions have rate 1. The blue arrows denote the reaction $X+{\rm ATP} \rightleftharpoons X^* + {\rm ADP}$.}
 \label{fig:summary}
\end{figure}

As illustrated in Fig.~\ref{fig:summary}, we pull a polymer decorated with substrates $X$ and $X^*$ past an enzyme tethered within a nucleotide bath. The regularly-spaced substrates constitute ``tape" $A$, with sites $A_n$. We shall discuss the second tape $B$, and operational details, later. The exploitable free energy per monomer stored in tape $A$ alone depends on the intrinsic chemical potentials of $X$ and $X^*$, $\mu^0_X$ and $\mu^0_{X^*}$, and the initial fraction of $X^*$,  $q=p_0(A_n = X^*)$. Specifically, assuming  $A_n$ are independent,
$
\tilde{F}(A_n) - F_{\rm eq}(A_n) = (q-q_{\rm eq})(\mu^0_{X^*} - \mu^0_X) + k_BT (H_{\rm eq}(A_n)- H(A_n))
$,
where $q_{\rm eq}$ is the equilibrium fraction of $X^*$ when the substrates are decoupled from ATP and ADP, $q_{\rm eq}=p_{\rm eq}(A_n = X^*) = (1+ \exp((\mu^0_{X^*} - \mu^0_X)/k_BT))^{-1} $. If left uncoupled from ATP and ADP, $A_n$ would eventually reach this equilibrium due to  spontaneous transitions $X + {\rm {P_i}}\leftrightarrow X^{*}$ that we assume are slow enough to be ignored under machine operation.

An active enzyme, supplied only with tape $A$, could do work on the bath if $\tilde{F}(A_n) - F_{\rm eq}(A_n) >0$, implying an excess of the most unstable substrate, a low entropy $H(A_n)$, or both. If $\mu^0_X = \mu^0_{X^*}$ this $A$-only system is equivalent to the device of Ref. \cite{Mandal2012}, with $X/X^*$ as degenerate 0s and 1s storing free energy purely through $H(A_n)< H_{\rm eq}(A_n)=\ln 2$. Cao {\it et al.}  \cite{Cao2015} proposed using an enzyme in this context, but without a physical tape. In these studies it appears remarkable that abstract strings of bits can be used for work. In a physical implementation, however, the devices are powered by non-equilibrium fuel, like all machines. For general $\mu^0_X - \mu^0_{X^*}$, equilibrium states have  $H_{\rm eq}(A_n)< \ln 2$. One cannot, therefore, infer that  tape $A$ is a fuel source from   $H(A_n) $  alone; $\mu^0_X - \mu^0_{X^*}$  must  be considered.

Our system has a second tape ``$B$" grafted onto the polymer, moving together with $A$ (Fig.~\ref{fig:summary}). Each $n$ thus corresponds to a pair $A_n$ and $B_n$, with $B_n$ carrying $Y$ or an empty space $\bar{Y}$. The inter-site interval is long enough that $E$ can interact with at most one pair simultaneously, and $A_n, B_n$ are initially independent of  $A_m, B_m$ for $n \neq m$. Operational cycles begin with  $E$ isolated from the sites on either tape, and in equilibrium with respect to ATP/ADP binding. We then bring the next pair of sites, $n$, alongside $E$. If $B_n=Y$, $E \rightarrow E^\dagger$ and conversion of $A_n$ between $X$ and $X^{*}$ according to Eq.~\ref{eq:chem1} is possible ($B_n$ remains unchanged). We then move the tapes along,  allowing $E$ to relax back into equilibrium with the chemical bath. We return to the costs of external manipulation later. Tape $B$ provides an additional positive contribution to the exploitable free energy, $k_BTI(A_n;B_n)$ (Eq.\,\ref{eq:info}). Physically, the enzyme's decision-making interaction with $B_n$ sets its response to $A_n$, since $A_n$ is fixed when $B_n=\bar{Y}$. If $Y$ is  correlated with $X^{*}$, then $X^{*} + {\rm ADP} \rightarrow X + {\rm ATP}$ happens with increased frequency relative to
$X + {\rm ATP} \rightarrow X^* + {\rm ADP}$, {\em allowing work extraction  even if} $\tilde{F}(A_n) - F_{\rm eq}(A_n) =0$. Unlike $H(A_n) < \ln 2$,  $I(A_n;B_n) \neq 0$ always implies stored free energy.

A proposal for instantiating the system, using DNA origami assembly \cite{Rothemund06,Douglas09}, is given in Section 1 of Ref.\,\cite{SI}. Here we analyse a simple model of device  operation. Recall that $p_0(A_n = X^*)=q$. For simplicity we introduce  initial correlations through a single parameter $\psi$:
\begin{align*}
p_0(B_{n} = \bar{Y} | A_{n} &= X^{*}) = p_0(B_{n} = Y | A_{n} = X) = \psi, \numberthis \label{eq:psi_def}\\
p_0(B_{n} = Y | A_{n} &= X^{*}) = p_0(B_{n} = \bar{Y} | A_{n} = X) = 1-\psi.
\end{align*}
We adjust the chemical load from the bath via  $\alpha \in (-1, 1)$ such that $[{\rm ATP}] = 1+\alpha$ and $[{\rm ADP}] = 1-\alpha$ relative to a reference concentration $C_0$. 

\begin{figure*}
\includegraphics[width=\textwidth]{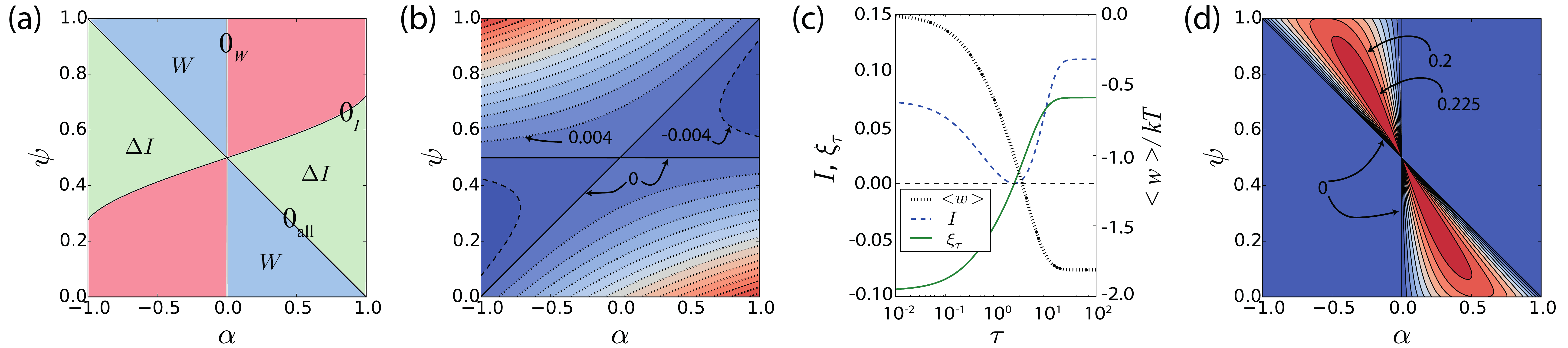}
\caption{(a) Regimes of operation for  $q=0.5$, $\tau\to\infty$ as a function of chemical load $\alpha$ and correlation strength $\psi$ (see Eq.~\ref{eq:psi_def}): $W$ indicates $\langle w_{\rm cycle} \rangle > 0$, $\Delta I$ indicates $\Delta I(A_n;B_n) > 0$; $\langle w_{\rm cycle} \rangle$ and $\Delta I(A_n;B_n)$ are negative elsewhere. (b) Product of covariances $\xi_0(A_n,B_n) \xi_\tau(A_n,B_n)$ for $q=0.5$, $\tau\to\infty$, showing correlation reversal. Covariances calculated by taking $X^*$ and $Y$ as 1, and $X$ and $\bar{Y}$ as 0.  Negative contours are dashed lines, positive are dotted and zero is solid (contours separated by units of 0.004). (c) Behaviour of $\langle w_{\rm cycle}\rangle$, $\Delta I(A_n;B_n)$ and correlation  $\xi_\tau(A_n,B_n)$ as a function of $\tau$ for $q=0.5$, $\alpha=0.99$ and $\psi=0.69$, showing non-monotonicity of information. (d) Efficiency $\eta$ of converting initial mutual information into work. Contours run downwards from 0.225 with a separation of 0.025.  \label{fig:q0.5}}
\end{figure*}

We model tape manipulation  as exposing $A_n$ to the enzyme for a time $\tau$, assuming that allostery and activator binding are sufficiently strong that the enzyme is  active  if $Y$ is present, but inactive otherwise. Thus the relevant transitions are binding/unbinding \textcolor{black}{of nucleotides} to $E$, and the catalysis of Eq.~\ref{eq:chem1}. Enzyme/substrate unbinding is assumed \textcolor{black}{fast, and catalysis instantaneous}. For simplicity, we set all rate constants to unity (1/$C_0$ for bimolecular rate constants), and assume that $\mu_{\rm ADP} - \mu_{\rm ATP} = kT \ln \frac{1-\alpha}{1+\alpha}$, implying that \textcolor{black}{$\mu^0_X=\mu^0_{X^*}$} given these rate constants. We relax these assumptions in Ref.\,\cite{SI}, Section 3. 
The model has the transition graph shown in Fig.~\ref{fig:summary}\,c, which \textcolor{black}{specifies a reaction rate matrix $\mathcal{R}$ and probability density evolution} $p_{t} = {\rm e}^{\mathcal{R}t}p_{0}$. Since the \textcolor{black}{ADP/ATP-binding state of $E$ equilibrates between encounters with  sites}, and pairs of sites are initially independent, we can consider each pair  $A_n$, $B_n$ separately. See Ref.\,\cite{SI}, Section 2, for full dynamical solutions.

For $\tau \rightarrow \infty$, substrates paired with $B_n =Y$ relax to \textcolor{black}{a new} equilibrium with the bath. 
If $\alpha$ is positive, this equilibrium favours an excess of $X^*$ since $[{\rm ATP}] > [{\rm ADP}]$. The device can only do work on the bath if the  substrates encountered by active enzymes are even more likely to be in state $X^*$ than this excess.
For $q=0.5$ and $\mu^0_X=\mu^0_{X^*}$, tape $A$ is initially in  equilibrium, $\tilde{F}_0(A_n) = {F}_{\rm eq}(A_n)$, and thus cannot be used to do work \textcolor{black}{without harnessing $I_0(A_n;B_n)$}. Correlations allow  $E^\dagger$  to encounter  an excess of $X^*$ ($X$) if $\psi<0.5$ ($\psi > 0.5$), despite $q=0.5$. See Ref.\,\cite{SI}, Section 2 for the resultant work and change in mutual information as a function of $\alpha$ and $\psi$. 

\textcolor{black}{We indicate operational regimes for $q=0.5$ in Fig.~\ref{fig:q0.5}\,a, using $\Delta$ to denote the change of a quantity during $\tau$. In the bottom right, $\Delta I(A_n;B_n)<0$ and $\langle w_{\rm cycle} \rangle >0$: the correlations favour $X^{*} + {\rm ADP} \rightarrow X + {\rm ATP}$ and are strong enough to do work against the load $\mu_{\rm ATP} - \mu_{\rm ADP}$. This region is bounded by the loci $0_W$, along which $\langle w_{\rm cycle} \rangle =0$ since the load is zero ($\mu_{\rm ATP} = \mu_{\rm ADP}$), and $0_{\rm all}$ ($\alpha = 1 - 2\psi$,  Ref.~\cite{SI}, section 2), along which the correlations and  load exactly balance, implying no net evolution and  $\Delta I(A_n;B_n) = \langle w_{\rm cycle} \rangle/k_{B}T=0$. Directly above is a region with $\Delta I(A_n;B_n)>0$ and $\langle w_{\rm cycle} \rangle <0$, in which the chemical load is strong enough to do work on the tapes, creating information. Finally, above this regime, both $\Delta I(A_n;B_n)<0$ and $\langle w_{\rm cycle} \rangle <0$. These regimes are repeated on the left hand side of the figure.}

 The final two regimes are divided by the locus \textcolor{black}{$0_{I}$, along which $\Delta I(A_n;B_n) = 0$. The dynamics here is clarified by considering the covariance $\xi$ between $A_n$ and $B_n$, which quantifies not only  correlation strength but also whether $X^{*}$ is associated with $Y$ or $\bar{Y}$. Along $0_{I}$, $X^{*}$ typically appears with $\bar{Y}$ for $\alpha > 0$ and with $Y$ for $\alpha < 0$. Thus both the load and correlations initially favour $X^{*}\rightarrow X$ for $\alpha < 0$ and $X \rightarrow X^{*}$ for $\alpha > 0$. As the reactions proceed the correlations between $A_{n}$ and $B_{n}$ drop and eventually reverse as substrates are converted. The resultant excess of the alternative substrate opposes the chemical load and eventually halts the reaction at $\Delta I(A_n;B_n)=0$, but with inverted correlations (Fig.~\ref{fig:q0.5}\,b).}

Considering finite $\tau$  highlights correlation inversion; see Fig.~\ref{fig:q0.5}\,c for $\alpha=0.99$ and $\psi=0.69$, parameters close to the \textcolor{black}{$0_I$ locus at $\tau \rightarrow \infty$ (shifting the $0_{I}$ locus is the main effect of finite $\tau$,} see Ref.\,\cite{SI}, Section 3). The reversal of correlations with $\tau$ demonstrates \textcolor{black}{the difference between each pair of sites $A_n$, $B_n$} and the chemical bath; they function like a charged capacitor and a constant voltage supply, respectively. If connected in series, a capacitor and supply can initially work together; eventually, however, the capacitor discharges and then recharges with the opposite polarity. 

We plot the fraction of $I_0(A_n,B_n)$ converted into work, $\eta=\langle w_{\rm cycle} \rangle/k_B T I_0(A_n;B_n)$, in Fig.~\ref{fig:q0.5}\,d. The maximum $\eta$ is $\sim1/4$, midway between $0_{\rm all}$ and $0_{W}$. Near $0_{\rm all}$,  the system \textcolor{black}{is near} equilibrium and the evolution involves negligible dissipation, but also very little actual change during $\tau$, so  $\Delta I(A_n;B_n) \approx 0$. Near $0_{W}$, the system evolves significantly during $\tau$, but does so wastefully as the chemical load is too small. See Ref.\,\cite{SI}, Section 2, for further analysis of efficiency.

\begin{figure}
\includegraphics[width=0.49\textwidth]{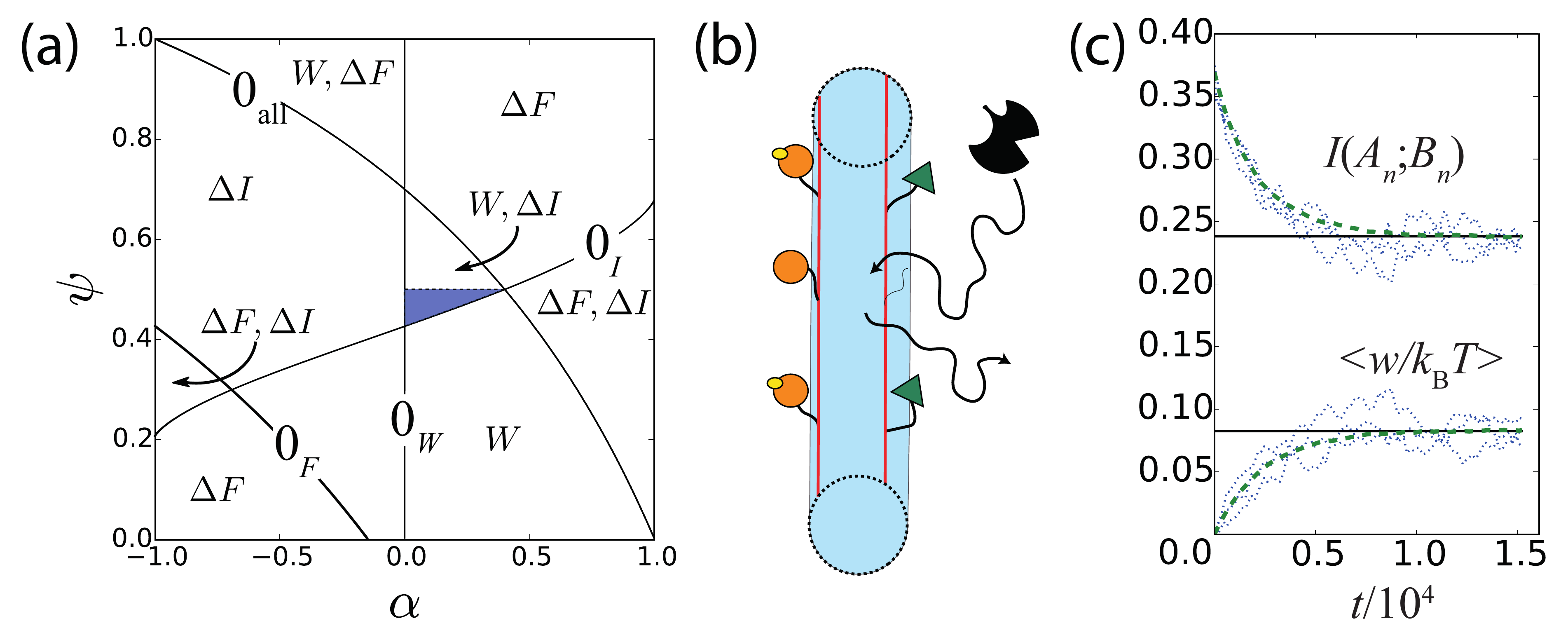}
\caption{(a) Regimes of behaviour for $q=0.7$, $\tau \rightarrow \infty$: $\Delta F$ indicates  $\Delta \tilde{F}(A_n) > 0$; $W$ and $\Delta I$ are defined equivalently. Highlighted (blue) region has $\Delta I(A_n;B_n) \geq 0$ and a greater  $\langle w_{\rm cycle} \rangle$ than for a system with the same $q,\alpha$, but $I_0(A_n;B_n)=0$. (b) A completely autonomous device in  which a freely-diffusing enzyme interacts with correlated tapes. (c) Autonomous dynamics, with $q=0.5$, $\psi=0.1$ and $\alpha=0.5$. In addition to the rate constants in Fig.\,\ref{fig:summary}\,c, it is necessary to define a rate at which the enzyme binds to activator $Y$ ($k_{\rm bind} = 0.01$ per activator here) and an unbinding rate ($k_{\rm off} = 1$ here). We plot the average work per site $\langle w \rangle/kT$ and $I(A_n;B_n)$ obtained from 200 independent Gillespie simulations of 1000 pairs of sites \cite{Gillespie1977} (dashed lines), showing that they converge to the values predicted in the $\tau \rightarrow \infty$ limit of the externally controlled device (solid lines). Dotted lines show information and work within three individual simulations (for subtleties, see Ref.\,\cite{SI}, Section 6).   \label{fig:free-running}}
\end{figure}

For $q\neq0.5$,  $\tilde{F}_0(A_n) \neq {F}_{\rm eq}(A_n)$,  and tape $A$ \textcolor{black}{itself stores free energy}. New regimes of behaviour arise, shown in Fig.~\ref{fig:free-running}\,a for $q=0.7$. We see all sign permutations of $\Delta \tilde{F}(A_n)$, $\Delta I(A_n;B_n)$ and $\langle w_{\rm cycle}\rangle$ except for the second law-violating $\Delta \tilde{F}(A_n),\,\Delta I(A_n;B_n),\,\langle w_{\rm cycle}\rangle>0$. The initial value of $\tilde{F}(A_n) \neq F_{\rm eq} (A_n)$ distorts  regimes and introduces a new boundary, \textcolor{black}{$0_F$}. Locus \textcolor{black}{$0_F$} is analogous to \textcolor{black}{$0_I$}: along \textcolor{black}{$0_F$}, the probabilities of $X^*$ and $X$ are inverted by  machine operation, preserving $\tilde{F}(A_n) =\tilde{F}_0(A_n)  $.

\textcolor{black}{When tape $A$ itself is a non-equilibrium fuel}, non-zero $I_0(A_n;B_n)$ allows for {\it increased} $\langle w_{\rm cycle}\rangle$ in two ways. The most intuitive is analogous to the $q=0.5$ case: $I(A_n;B_n)$ is consumed to drive catalysis against the bath's chemical load. More surprisingly, in the highlighted region of Fig.~\ref{fig:free-running}\,a, $\Delta I(A_n;B_n) \geq 0$ yet $\langle w_{\rm cycle}\rangle$ is increased relative to an otherwise equivalent system with $I_0(A_n; B_n)=0$. Here, correlations initially support the drive from $\tilde{F}(A_n)\neq  {F}_{\rm eq}(A_n)$ to do work against the bath, but eventually reverse and oppose further reactions. The need to first reverse the initial correlations allows more ATP production  before the reaction halts, enhancing $\langle w_{\rm cycle} \rangle$ despite $\Delta I(A_n;B_n) \geq 0$.

Each trajectory $z(t)=(e(t),a_n(t),b_n(t))$, with $e(t)$, $a_n(t)$ and $b_n(t)$ representing the states of $E$, $A_n$ and $B_n$ over time, \textcolor{black}{produces} a stochastic entropy $\sigma[z(t),p_t(z)]$,  a functional of $z(t)$ and the probability of occupying state $z$ at time $t$,  $p_t(z)$  \cite{Crooks1999,Seifert2011,Seifert2012}. With $b_n(t)$ constant, the fluctuation relation 
$\langle {\rm e}^{-\bar{\sigma}[e_n(t),a_n(t),{p}_t(e,a_n)]/k_B + \Delta i_{0,\tau}(b_n;e,a_n)}\rangle = 1$ holds,
where $\bar{\sigma}$ is the stochastic entropy generated by the $(E, A_n)$ subsystem and $i(b_n;e,a_n) = \ln(p(b_n,e,a_n)/p(b_n)p(e,a_n))$ is a pointwise mutual information \cite{Sagawa2012}.
\textcolor{black}{This result shows  at the trajectory level how consumption of 
 mutual information compensates for entropy reduction elsewhere, permitting} $\Delta F(A_n)>0$ or $\langle w_{\rm cycle}\rangle>0$. Unusually, in our model the fluctuation theorem is  verifiable by explicit summation  (Ref.\,\cite{SI}, Section 4).

 Fig.\,\ref{fig:summary}\,a,b emphasises that the information register physically couples to the device. If the tapes are pulled at a finite rate, irreversible work will be done on the protein-protein interactions. \textcolor{black}{Minimising this dissipation through slow tape manipulation is problematic due to unwanted spontaneous transitions $X + {\rm {P_i}}\leftrightarrow X^{*}$}. Additional subtleties also arise when unidirectional motion is imposed externally \cite{Machta2015}, as outlined in Ref.\,\cite{SI}, Section 5.
These problems are eliminated if the device is autonomous, with no external manipulation after initiation.
We therefore propose a modified autonomous system with an enzyme that   \textcolor{black}{diffuses freely between static sites} (Fig.\,\ref{fig:free-running}\,b). The resultant dynamics are more complex -- the enzyme can repeatedly return to the same  sites -- but  $I_0(A_n;B_n)$ still allows selective interaction with either $X$ or $X^*$. Indeed, autonomous systems in the long time limit and non-autonomous systems with $\tau \rightarrow \infty$ perform identically. In both cases, the enzyme  allows those substrates paired with $Y$ to relax to equilibrium with the bath, and thus the limiting distribution is the same for a given set of $\alpha$, $\psi$ and $q$. Thus, as shown in Fig.\,\ref{fig:free-running}\,c, the long-time average behaviour of the autonomous system reproduces the previously reported behaviour. 

\textcolor{black}{Many sophisticated behaviours  exploit correlations in the environment, and we have taken the first step in designing minimal artificial systems that do this in the most fundamental way possible, by extracting work. Like biological systems, our second design can function autonomously. The chemical free energy stored  could power a molecular motor, for example, and an eventual goal would be to design a minimal system that actually used the harvested free energy to support itself. Our devices lack memory, since the information exploited is the correlation between degrees of freedom that are encountered simultaneously. Designing systems that can harness correlations over time is the subject of future work.}

\bibliography{two_tape_bib}

\begin{thebibliography}{45}%
\makeatletter
\providecommand \@ifxundefined [1]{%
 \@ifx{#1\undefined}
}%
\providecommand \@ifnum [1]{%
 \ifnum #1\expandafter \@firstoftwo
 \else \expandafter \@secondoftwo
 \fi
}%
\providecommand \@ifx [1]{%
 \ifx #1\expandafter \@firstoftwo
 \else \expandafter \@secondoftwo
 \fi
}%
\providecommand \natexlab [1]{#1}%
\providecommand \enquote  [1]{``#1''}%
\providecommand \bibnamefont  [1]{#1}%
\providecommand \bibfnamefont [1]{#1}%
\providecommand \citenamefont [1]{#1}%
\providecommand \href@noop [0]{\@secondoftwo}%
\providecommand \href [0]{\begingroup \@sanitize@url \@href}%
\providecommand \@href[1]{\@@startlink{#1}\@@href}%
\providecommand \@@href[1]{\endgroup#1\@@endlink}%
\providecommand \@sanitize@url [0]{\catcode `\\12\catcode `\$12\catcode
  `\&12\catcode `\#12\catcode `\^12\catcode `\_12\catcode `\%12\relax}%
\providecommand \@@startlink[1]{}%
\providecommand \@@endlink[0]{}%
\providecommand \url  [0]{\begingroup\@sanitize@url \@url }%
\providecommand \@url [1]{\endgroup\@href {#1}{\urlprefix }}%
\providecommand \urlprefix  [0]{URL }%
\providecommand \Eprint [0]{\href }%
\providecommand \doibase [0]{http://dx.doi.org/}%
\providecommand \selectlanguage [0]{\@gobble}%
\providecommand \bibinfo  [0]{\@secondoftwo}%
\providecommand \bibfield  [0]{\@secondoftwo}%
\providecommand \translation [1]{[#1]}%
\providecommand \BibitemOpen [0]{}%
\providecommand \bibitemStop [0]{}%
\providecommand \bibitemNoStop [0]{.\EOS\space}%
\providecommand \EOS [0]{\spacefactor3000\relax}%
\providecommand \BibitemShut  [1]{\csname bibitem#1\endcsname}%
\let\auto@bib@innerbib\@empty
\bibitem [{\citenamefont {Micali}\ and\ \citenamefont
  {Endres}(2016)}]{Micali2016}%
  \BibitemOpen
  \bibfield  {author} {\bibinfo {author} {\bibfnamefont {G.}~\bibnamefont
  {Micali}}\ and\ \bibinfo {author} {\bibfnamefont {R.~G.}\ \bibnamefont
  {Endres}},\ }\href@noop {} {\bibfield  {journal} {\bibinfo  {journal} {Curr.
  Opin. Microbiol.}\ }\textbf {\bibinfo {volume} {30}},\ \bibinfo {pages} {8}
  (\bibinfo {year} {2016})}\BibitemShut {NoStop}%
\bibitem [{\citenamefont {Becker}\ \emph {et~al.}(2015)\citenamefont {Becker},
  \citenamefont {Mugler},\ and\ \citenamefont {ten Wolde}}]{Becker2015}%
  \BibitemOpen
  \bibfield  {author} {\bibinfo {author} {\bibfnamefont {N.~B.}\ \bibnamefont
  {Becker}}, \bibinfo {author} {\bibfnamefont {A.}~\bibnamefont {Mugler}}, \
  and\ \bibinfo {author} {\bibfnamefont {P.~R.}\ \bibnamefont {ten Wolde}},\
  }\href@noop {} {\bibfield  {journal} {\bibinfo  {journal} {Phys. Rev. Lett.}\
  }\textbf {\bibinfo {volume} {115}},\ \bibinfo {pages} {258103} (\bibinfo
  {year} {2015})}\BibitemShut {NoStop}%
\bibitem [{\citenamefont {Shannon}\ and\ \citenamefont
  {Weaver}(1949)}]{Shannon1949}%
  \BibitemOpen
  \bibfield  {author} {\bibinfo {author} {\bibfnamefont {C.~E.}\ \bibnamefont
  {Shannon}}\ and\ \bibinfo {author} {\bibfnamefont {W.}~\bibnamefont
  {Weaver}},\ }\href@noop {} {\emph {\bibinfo {title} {The mathematical theory
  of communication}}}\ (\bibinfo  {publisher} {University of Illinois Press},\
  \bibinfo {year} {1949})\BibitemShut {NoStop}%
\bibitem [{\citenamefont {Govern}\ and\ \citenamefont {ten
  Wolde}(2014)}]{Govern2014}%
  \BibitemOpen
  \bibfield  {author} {\bibinfo {author} {\bibfnamefont {C.~C.}\ \bibnamefont
  {Govern}}\ and\ \bibinfo {author} {\bibfnamefont {P.~R.}\ \bibnamefont {ten
  Wolde}},\ }\href {\doibase 10.1103/PhysRevLett.113.258102} {\bibfield
  {journal} {\bibinfo  {journal} {Phys. Rev. Lett.}\ }\textbf {\bibinfo
  {volume} {113}},\ \bibinfo {pages} {258102} (\bibinfo {year}
  {2014})}\BibitemShut {NoStop}%
\bibitem [{\citenamefont {Barato}\ \emph {et~al.}(2014)\citenamefont {Barato},
  \citenamefont {Hartich},\ and\ \citenamefont {Seifert}}]{Barato2014}%
  \BibitemOpen
  \bibfield  {author} {\bibinfo {author} {\bibfnamefont {A.~C.}\ \bibnamefont
  {Barato}}, \bibinfo {author} {\bibfnamefont {D.}~\bibnamefont {Hartich}}, \
  and\ \bibinfo {author} {\bibfnamefont {U.}~\bibnamefont {Seifert}},\
  }\href@noop {} {\bibfield  {journal} {\bibinfo  {journal} {New. J. Phys.}\
  }\textbf {\bibinfo {volume} {16}},\ \bibinfo {pages} {103024} (\bibinfo
  {year} {2014})}\BibitemShut {NoStop}%
\bibitem [{\citenamefont {Ouldridge}\ \emph {et~al.}()\citenamefont
  {Ouldridge}, \citenamefont {Govern},\ and\ \citenamefont {ten
  Wolde}}]{Ouldridge2015}%
  \BibitemOpen
  \bibfield  {author} {\bibinfo {author} {\bibfnamefont {T.~E.}\ \bibnamefont
  {Ouldridge}}, \bibinfo {author} {\bibfnamefont {C.~C.}\ \bibnamefont
  {Govern}}, \ and\ \bibinfo {author} {\bibfnamefont {P.~R.}\ \bibnamefont {ten
  Wolde}},\ }\href@noop {} {\bibinfo  {journal} {arXiv:1503.00909}\
  }\BibitemShut {NoStop}%
\bibitem [{\citenamefont {Cheong}\ \emph {et~al.}(2011)\citenamefont {Cheong},
  \citenamefont {Rhee}, \citenamefont {Wang}, \citenamefont {Nemenman},\ and\
  \citenamefont {Levchenko}}]{Cheong2011}%
  \BibitemOpen
\bibfield  {journal} {  }\bibfield  {author} {\bibinfo {author} {\bibfnamefont
  {R.}~\bibnamefont {Cheong}}, \bibinfo {author} {\bibfnamefont
  {A.}~\bibnamefont {Rhee}}, \bibinfo {author} {\bibfnamefont {C.~J.}\
  \bibnamefont {Wang}}, \bibinfo {author} {\bibfnamefont {I.}~\bibnamefont
  {Nemenman}}, \ and\ \bibinfo {author} {\bibfnamefont {A.}~\bibnamefont
  {Levchenko}},\ }\href@noop {} {\bibfield  {journal} {\bibinfo  {journal}
  {Science}\ }\textbf {\bibinfo {volume} {334}},\ \bibinfo {pages} {354}
  (\bibinfo {year} {2011})}\BibitemShut {NoStop}%
\bibitem [{\citenamefont {de~Ronde}\ \emph {et~al.}(2011)\citenamefont
  {de~Ronde}, \citenamefont {Tostevin},\ and\ \citenamefont {ten
  Wolde}}]{deRonde2011}%
  \BibitemOpen
  \bibfield  {author} {\bibinfo {author} {\bibfnamefont {W.}~\bibnamefont
  {de~Ronde}}, \bibinfo {author} {\bibfnamefont {F.}~\bibnamefont {Tostevin}},
  \ and\ \bibinfo {author} {\bibfnamefont {P.~R.}\ \bibnamefont {ten Wolde}},\
  }\href@noop {} {\bibfield  {journal} {\bibinfo  {journal} {Phys. Rev. Lett.}\
  }\textbf {\bibinfo {volume} {107}},\ \bibinfo {pages} {048101} (\bibinfo
  {year} {2011})}\BibitemShut {NoStop}%
\bibitem [{\citenamefont {Sartori}\ \emph {et~al.}(2014)\citenamefont
  {Sartori}, \citenamefont {Granger}, \citenamefont {Lee},\ and\ \citenamefont
  {Horowitz}}]{Sartori2014}%
  \BibitemOpen
  \bibfield  {author} {\bibinfo {author} {\bibfnamefont {P.}~\bibnamefont
  {Sartori}}, \bibinfo {author} {\bibfnamefont {L.}~\bibnamefont {Granger}},
  \bibinfo {author} {\bibfnamefont {C.~F.}\ \bibnamefont {Lee}}, \ and\
  \bibinfo {author} {\bibfnamefont {J.~M.}\ \bibnamefont {Horowitz}},\ }\href
  {\doibase 10.1371/journal.pcbi.1003974} {\bibfield  {journal} {\bibinfo
  {journal} {PLoS Comput Biol}\ }\textbf {\bibinfo {volume} {10}},\ \bibinfo
  {pages} {e1003974} (\bibinfo {year} {2014})}\BibitemShut {NoStop}%
\bibitem [{\citenamefont {Ito}\ and\ \citenamefont {Sagawa}(2015)}]{Ito2014}%
  \BibitemOpen
  \bibfield  {author} {\bibinfo {author} {\bibfnamefont {S.}~\bibnamefont
  {Ito}}\ and\ \bibinfo {author} {\bibfnamefont {T.}~\bibnamefont {Sagawa}},\
  }\href@noop {} {\bibfield  {journal} {\bibinfo  {journal} {Nat. Comm.}\
  }\textbf {\bibinfo {volume} {6}},\ \bibinfo {pages} {7498} (\bibinfo {year}
  {2015})}\BibitemShut {NoStop}%
\bibitem [{\citenamefont {Bialek}(2015)}]{Bialek2015}%
  \BibitemOpen
  \bibfield  {author} {\bibinfo {author} {\bibfnamefont {W.}~\bibnamefont
  {Bialek}},\ }\href@noop {} {\bibfield  {journal} {\bibinfo  {journal}
  {arXiv:1512.08954}\ } (\bibinfo {year} {2015})}\BibitemShut {NoStop}%
\bibitem [{\citenamefont {Landauer}(1961)}]{Landauer1961}%
  \BibitemOpen
  \bibfield  {author} {\bibinfo {author} {\bibfnamefont {R.}~\bibnamefont
  {Landauer}},\ }\href@noop {} {\bibfield  {journal} {\bibinfo  {journal} {IBM
  J. Res. Dev.}\ }\textbf {\bibinfo {volume} {5}},\ \bibinfo {pages} {183}
  (\bibinfo {year} {1961})}\BibitemShut {NoStop}%
\bibitem [{\citenamefont {Bennett}(1982)}]{Bennett1982}%
  \BibitemOpen
  \bibfield  {author} {\bibinfo {author} {\bibfnamefont {C.~H.}\ \bibnamefont
  {Bennett}},\ }\href@noop {} {\bibfield  {journal} {\bibinfo  {journal} {Int.
  J. Theor. Phys.}\ }\textbf {\bibinfo {volume} {21}},\ \bibinfo {pages} {905}
  (\bibinfo {year} {1982})}\BibitemShut {NoStop}%
\bibitem [{\citenamefont {Sagawa}\ and\ \citenamefont
  {Ueda}(2009)}]{Sagawa2009}%
  \BibitemOpen
  \bibfield  {author} {\bibinfo {author} {\bibfnamefont {T.}~\bibnamefont
  {Sagawa}}\ and\ \bibinfo {author} {\bibfnamefont {M.}~\bibnamefont {Ueda}},\
  }\href@noop {} {\bibfield  {journal} {\bibinfo  {journal} {Phys. Rev. Lett.}\
  }\textbf {\bibinfo {volume} {102}},\ \bibinfo {pages} {250602} (\bibinfo
  {year} {2009})}\BibitemShut {NoStop}%
\bibitem [{\citenamefont {Bauer}\ \emph {et~al.}(2012)\citenamefont {Bauer},
  \citenamefont {Abreu},\ and\ \citenamefont {Seifert}}]{Bauer2012}%
  \BibitemOpen
  \bibfield  {author} {\bibinfo {author} {\bibfnamefont {M.}~\bibnamefont
  {Bauer}}, \bibinfo {author} {\bibfnamefont {D.}~\bibnamefont {Abreu}}, \ and\
  \bibinfo {author} {\bibfnamefont {U.}~\bibnamefont {Seifert}},\ }\href@noop
  {} {\bibfield  {journal} {\bibinfo  {journal} {J. Phys. A-Math. Theor.}\
  }\textbf {\bibinfo {volume} {45}},\ \bibinfo {pages} {162001} (\bibinfo
  {year} {2012})}\BibitemShut {NoStop}%
\bibitem [{\citenamefont {Still}\ \emph {et~al.}(2012)\citenamefont {Still},
  \citenamefont {Sivak}, \citenamefont {Bell},\ and\ \citenamefont
  {Crooks}}]{Still2012}%
  \BibitemOpen
  \bibfield  {author} {\bibinfo {author} {\bibfnamefont {S.}~\bibnamefont
  {Still}}, \bibinfo {author} {\bibfnamefont {D.~A.}\ \bibnamefont {Sivak}},
  \bibinfo {author} {\bibfnamefont {A.~J.}\ \bibnamefont {Bell}}, \ and\
  \bibinfo {author} {\bibfnamefont {G.~E.}\ \bibnamefont {Crooks}},\
  }\href@noop {} {\bibfield  {journal} {\bibinfo  {journal} {Phys. Rev. Lett.}\
  }\textbf {\bibinfo {volume} {109}},\ \bibinfo {pages} {120604} (\bibinfo
  {year} {2012})}\BibitemShut {NoStop}%
\bibitem [{\citenamefont {Sagawa}\ and\ \citenamefont
  {Ueda}(2012)}]{Sagawa2012}%
  \BibitemOpen
  \bibfield  {author} {\bibinfo {author} {\bibfnamefont {T.}~\bibnamefont
  {Sagawa}}\ and\ \bibinfo {author} {\bibfnamefont {M.}~\bibnamefont {Ueda}},\
  }\href@noop {} {\bibfield  {journal} {\bibinfo  {journal} {Phys. Rev. Lett.}\
  }\textbf {\bibinfo {volume} {109}},\ \bibinfo {pages} {180602} (\bibinfo
  {year} {2012})}\BibitemShut {NoStop}%
\bibitem [{\citenamefont {Horowitz}\ and\ \citenamefont
  {Esposito}(2014)}]{Horowitz2014}%
  \BibitemOpen
  \bibfield  {author} {\bibinfo {author} {\bibfnamefont {J.~M.}\ \bibnamefont
  {Horowitz}}\ and\ \bibinfo {author} {\bibfnamefont {M.}~\bibnamefont
  {Esposito}},\ }\href@noop {} {\bibfield  {journal} {\bibinfo  {journal}
  {Phys. Rev. X}\ }\textbf {\bibinfo {volume} {4}},\ \bibinfo {pages} {031015}
  (\bibinfo {year} {2014})}\BibitemShut {NoStop}%
\bibitem [{\citenamefont {Barato}\ and\ \citenamefont
  {Seifert}(2014)}]{Barato2014b}%
  \BibitemOpen
  \bibfield  {author} {\bibinfo {author} {\bibfnamefont {A.~C.}\ \bibnamefont
  {Barato}}\ and\ \bibinfo {author} {\bibfnamefont {U.}~\bibnamefont
  {Seifert}},\ }\href@noop {} {\bibfield  {journal} {\bibinfo  {journal} {Phys.
  Rev. Lett.}\ }\textbf {\bibinfo {volume} {112}},\ \bibinfo {pages} {090601}
  (\bibinfo {year} {2014})}\BibitemShut {NoStop}%
\bibitem [{\citenamefont {Parrondo}\ \emph {et~al.}(2015)\citenamefont
  {Parrondo}, \citenamefont {Horowitz},\ and\ \citenamefont
  {Sagawa}}]{Parrondo2015}%
  \BibitemOpen
  \bibfield  {author} {\bibinfo {author} {\bibfnamefont {J.~M.~R.}\
  \bibnamefont {Parrondo}}, \bibinfo {author} {\bibfnamefont {J.~M.}\
  \bibnamefont {Horowitz}}, \ and\ \bibinfo {author} {\bibfnamefont
  {T.}~\bibnamefont {Sagawa}},\ }\href@noop {} {\bibfield  {journal} {\bibinfo
  {journal} {Nat Phys}\ }\textbf {\bibinfo {volume} {11}},\ \bibinfo {pages}
  {131} (\bibinfo {year} {2015})}\BibitemShut {NoStop}%
\bibitem [{\citenamefont {Fahn}(1996)}]{Fahn1996}%
  \BibitemOpen
  \bibfield  {author} {\bibinfo {author} {\bibfnamefont {P.~N.}\ \bibnamefont
  {Fahn}},\ }\href@noop {} {\bibfield  {journal} {\bibinfo  {journal} {Found.
  Phys.}\ }\textbf {\bibinfo {volume} {26}},\ \bibinfo {pages} {71} (\bibinfo
  {year} {1996})}\BibitemShut {NoStop}%
\bibitem [{\citenamefont {Horowitz}\ \emph {et~al.}(2013)\citenamefont
  {Horowitz}, \citenamefont {Sagawa},\ and\ \citenamefont
  {Parrondo}}]{Horowitz2013}%
  \BibitemOpen
  \bibfield  {author} {\bibinfo {author} {\bibfnamefont {J.~M.}\ \bibnamefont
  {Horowitz}}, \bibinfo {author} {\bibfnamefont {T.}~\bibnamefont {Sagawa}}, \
  and\ \bibinfo {author} {\bibfnamefont {J.~M.~R.}\ \bibnamefont {Parrondo}},\
  }\href {\doibase 10.1103/PhysRevLett.111.010602} {\bibfield  {journal}
  {\bibinfo  {journal} {Phys. Rev. Lett.}\ }\textbf {\bibinfo {volume} {111}},\
  \bibinfo {pages} {010602} (\bibinfo {year} {2013})}\BibitemShut {NoStop}%
\bibitem [{\citenamefont {Esposito}\ and\ \citenamefont {Van
  Den~Broeck}(2011)}]{Esposito2011}%
  \BibitemOpen
  \bibfield  {author} {\bibinfo {author} {\bibfnamefont {M.}~\bibnamefont
  {Esposito}}\ and\ \bibinfo {author} {\bibfnamefont {C.}~\bibnamefont {Van
  Den~Broeck}},\ }\href@noop {} {\bibfield  {journal} {\bibinfo  {journal}
  {Europhys. Lett.}\ }\textbf {\bibinfo {volume} {95}},\ \bibinfo {pages}
  {40004} (\bibinfo {year} {2011})}\BibitemShut {NoStop}%
\bibitem [{\citenamefont {Mandal}\ and\ \citenamefont
  {Jarzynski}(2012)}]{Mandal2012}%
  \BibitemOpen
  \bibfield  {author} {\bibinfo {author} {\bibfnamefont {D.}~\bibnamefont
  {Mandal}}\ and\ \bibinfo {author} {\bibfnamefont {C.}~\bibnamefont
  {Jarzynski}},\ }\href@noop {} {\bibfield  {journal} {\bibinfo  {journal}
  {Proc. Nat. Acad. Sci. USA}\ }\textbf {\bibinfo {volume} {109}},\ \bibinfo
  {pages} {11641} (\bibinfo {year} {2012})}\BibitemShut {NoStop}%
\bibitem [{\citenamefont {Mandal}\ \emph {et~al.}(2013)\citenamefont {Mandal},
  \citenamefont {Quan},\ and\ \citenamefont {Jarzynski}}]{Mandal2013}%
  \BibitemOpen
  \bibfield  {author} {\bibinfo {author} {\bibfnamefont {D.}~\bibnamefont
  {Mandal}}, \bibinfo {author} {\bibfnamefont {H.~T.}\ \bibnamefont {Quan}}, \
  and\ \bibinfo {author} {\bibfnamefont {C.}~\bibnamefont {Jarzynski}},\
  }\href@noop {} {\bibfield  {journal} {\bibinfo  {journal} {Phys. Rev. Lett.}\
  }\textbf {\bibinfo {volume} {111}},\ \bibinfo {pages} {030602} (\bibinfo
  {year} {2013})}\BibitemShut {NoStop}%
\bibitem [{\citenamefont {Barato}\ and\ \citenamefont
  {Seifert}(2013)}]{Barato2013}%
  \BibitemOpen
  \bibfield  {author} {\bibinfo {author} {\bibfnamefont {A.~C.}\ \bibnamefont
  {Barato}}\ and\ \bibinfo {author} {\bibfnamefont {U.}~\bibnamefont
  {Seifert}},\ }\href {http://stacks.iop.org/0295-5075/101/i=6/a=60001}
  {\bibfield  {journal} {\bibinfo  {journal} {Europhys. Lett.}\ }\textbf
  {\bibinfo {volume} {101}},\ \bibinfo {pages} {60001} (\bibinfo {year}
  {2013})}\BibitemShut {NoStop}%
\bibitem [{\citenamefont {Lu}\ \emph {et~al.}(2014)\citenamefont {Lu},
  \citenamefont {Mandal},\ and\ \citenamefont {Jarzynski}}]{Lu2014}%
  \BibitemOpen
  \bibfield  {author} {\bibinfo {author} {\bibfnamefont {Z.}~\bibnamefont
  {Lu}}, \bibinfo {author} {\bibfnamefont {D.}~\bibnamefont {Mandal}}, \ and\
  \bibinfo {author} {\bibfnamefont {C.}~\bibnamefont {Jarzynski}},\ }\href@noop
  {} {\bibfield  {journal} {\bibinfo  {journal} {Phys. Today}\ }\textbf
  {\bibinfo {volume} {67}},\ \bibinfo {pages} {60} (\bibinfo {year}
  {2014})}\BibitemShut {NoStop}%
\bibitem [{\citenamefont {Cao}\ \emph {et~al.}(2015)\citenamefont {Cao},
  \citenamefont {Gong},\ and\ \citenamefont {Quan}}]{Cao2015}%
  \BibitemOpen
  \bibfield  {author} {\bibinfo {author} {\bibfnamefont {Y.}~\bibnamefont
  {Cao}}, \bibinfo {author} {\bibfnamefont {Z.}~\bibnamefont {Gong}}, \ and\
  \bibinfo {author} {\bibfnamefont {H.~T.}\ \bibnamefont {Quan}},\ }\href
  {\doibase 10.1103/PhysRevE.91.062117} {\bibfield  {journal} {\bibinfo
  {journal} {Phys. Rev. E}\ }\textbf {\bibinfo {volume} {91}},\ \bibinfo
  {pages} {062117} (\bibinfo {year} {2015})}\BibitemShut {NoStop}%
\bibitem [{\citenamefont {Boyd}\ \emph {et~al.}(2016)\citenamefont {Boyd},
  \citenamefont {Mandal},\ and\ \citenamefont {Crutchfield}}]{Boyd2016}%
  \BibitemOpen
  \bibfield  {author} {\bibinfo {author} {\bibfnamefont {A.~B.}\ \bibnamefont
  {Boyd}}, \bibinfo {author} {\bibfnamefont {D.}~\bibnamefont {Mandal}}, \ and\
  \bibinfo {author} {\bibfnamefont {J.~P.}\ \bibnamefont {Crutchfield}},\
  }\href@noop {} {\bibfield  {journal} {\bibinfo  {journal} {New J. Phys.}\
  }\textbf {\bibinfo {volume} {18}},\ \bibinfo {pages} {023049} (\bibinfo
  {year} {2016})}\BibitemShut {NoStop}%
\bibitem [{\citenamefont {Landauer}(1991)}]{Landauer1991}%
  \BibitemOpen
  \bibfield  {author} {\bibinfo {author} {\bibfnamefont {R.}~\bibnamefont
  {Landauer}},\ }\href@noop {} {\bibfield  {journal} {\bibinfo  {journal}
  {Phys. Today}\ }\textbf {\bibinfo {volume} {44}},\ \bibinfo {pages} {23}
  (\bibinfo {year} {1991})}\BibitemShut {NoStop}%
\bibitem [{\citenamefont {Toyabe}\ \emph {et~al.}(2010)\citenamefont {Toyabe},
  \citenamefont {Sagawa}, \citenamefont {Ueda}, \citenamefont {Muneyuki},\ and\
  \citenamefont {Sano}}]{toyabe2010experimental}%
  \BibitemOpen
  \bibfield  {author} {\bibinfo {author} {\bibfnamefont {S.}~\bibnamefont
  {Toyabe}}, \bibinfo {author} {\bibfnamefont {T.}~\bibnamefont {Sagawa}},
  \bibinfo {author} {\bibfnamefont {M.}~\bibnamefont {Ueda}}, \bibinfo {author}
  {\bibfnamefont {E.}~\bibnamefont {Muneyuki}}, \ and\ \bibinfo {author}
  {\bibfnamefont {M.}~\bibnamefont {Sano}},\ }\href@noop {} {\bibfield
  {journal} {\bibinfo  {journal} {Nat. Phys.}\ }\textbf {\bibinfo {volume}
  {6}},\ \bibinfo {pages} {988} (\bibinfo {year} {2010})}\BibitemShut {NoStop}%
\bibitem [{\citenamefont {Vidrighin}\ \emph {et~al.}(2016)\citenamefont
  {Vidrighin}, \citenamefont {Dahlsten}, \citenamefont {Barbieri},
  \citenamefont {Kim}, \citenamefont {Vedral},\ and\ \citenamefont
  {Walmsley}}]{Vidrighin2016}%
  \BibitemOpen
  \bibfield  {author} {\bibinfo {author} {\bibfnamefont {M.~D.}\ \bibnamefont
  {Vidrighin}}, \bibinfo {author} {\bibfnamefont {O.}~\bibnamefont {Dahlsten}},
  \bibinfo {author} {\bibfnamefont {M.}~\bibnamefont {Barbieri}}, \bibinfo
  {author} {\bibfnamefont {M.~S.}\ \bibnamefont {Kim}}, \bibinfo {author}
  {\bibfnamefont {V.}~\bibnamefont {Vedral}}, \ and\ \bibinfo {author}
  {\bibfnamefont {I.~A.}\ \bibnamefont {Walmsley}},\ }\href {\doibase
  10.1103/PhysRevLett.116.050401} {\bibfield  {journal} {\bibinfo  {journal}
  {Phys. Rev. Lett.}\ }\textbf {\bibinfo {volume} {116}},\ \bibinfo {pages}
  {050401} (\bibinfo {year} {2016})}\BibitemShut {NoStop}%
\bibitem [{\citenamefont {Chapman}\ and\ \citenamefont
  {Miyake}(2015)}]{Chapman2015}%
  \BibitemOpen
  \bibfield  {author} {\bibinfo {author} {\bibfnamefont {A.}~\bibnamefont
  {Chapman}}\ and\ \bibinfo {author} {\bibfnamefont {A.}~\bibnamefont
  {Miyake}},\ }\href {\doibase 10.1103/PhysRevE.92.062125} {\bibfield
  {journal} {\bibinfo  {journal} {Phys. Rev. E}\ }\textbf {\bibinfo {volume}
  {92}},\ \bibinfo {pages} {062125} (\bibinfo {year} {2015})}\BibitemShut
  {NoStop}%
\bibitem [{\citenamefont {Zeqiraj}\ \emph {et~al.}(2009)\citenamefont
  {Zeqiraj}, \citenamefont {Filippi}, \citenamefont {Deak}, \citenamefont
  {Alessi},\ and\ \citenamefont {van Aalten}}]{Zeqiraj2009}%
  \BibitemOpen
  \bibfield  {author} {\bibinfo {author} {\bibfnamefont {E.}~\bibnamefont
  {Zeqiraj}}, \bibinfo {author} {\bibfnamefont {B.~M.}\ \bibnamefont
  {Filippi}}, \bibinfo {author} {\bibfnamefont {M.}~\bibnamefont {Deak}},
  \bibinfo {author} {\bibfnamefont {D.~R.}\ \bibnamefont {Alessi}}, \ and\
  \bibinfo {author} {\bibfnamefont {D.~M.~F.}\ \bibnamefont {van Aalten}},\
  }\href@noop {} {\bibfield  {journal} {\bibinfo  {journal} {Science}\ }\textbf
  {\bibinfo {volume} {326}},\ \bibinfo {pages} {1707} (\bibinfo {year}
  {2009})}\BibitemShut {NoStop}%
\bibitem [{\citenamefont {Cowan-Jacob}\ \emph {et~al.}(2014)\citenamefont
  {Cowan-Jacob}, \citenamefont {Jahnke},\ and\ \citenamefont
  {Knapp}}]{Cowan-Jacob2014}%
  \BibitemOpen
  \bibfield  {author} {\bibinfo {author} {\bibfnamefont {S.~W.}\ \bibnamefont
  {Cowan-Jacob}}, \bibinfo {author} {\bibfnamefont {W.}~\bibnamefont {Jahnke}},
  \ and\ \bibinfo {author} {\bibfnamefont {S.}~\bibnamefont {Knapp}},\
  }\href@noop {} {\bibfield  {journal} {\bibinfo  {journal} {Future Med.
  Chem.}\ }\textbf {\bibinfo {volume} {6}},\ \bibinfo {pages} {541} (\bibinfo
  {year} {2014})}\BibitemShut {NoStop}%
\bibitem [{\citenamefont {Qiao}\ \emph {et~al.}(2006)\citenamefont {Qiao},
  \citenamefont {Molina}, \citenamefont {Pandey}, \citenamefont {Zhang},\ and\
  \citenamefont {Cole}}]{Qiao2006}%
  \BibitemOpen
  \bibfield  {author} {\bibinfo {author} {\bibfnamefont {Y.}~\bibnamefont
  {Qiao}}, \bibinfo {author} {\bibfnamefont {H.}~\bibnamefont {Molina}},
  \bibinfo {author} {\bibfnamefont {A.}~\bibnamefont {Pandey}}, \bibinfo
  {author} {\bibfnamefont {J.}~\bibnamefont {Zhang}}, \ and\ \bibinfo {author}
  {\bibfnamefont {P.~A.}\ \bibnamefont {Cole}},\ }\href@noop {} {\bibfield
  {journal} {\bibinfo  {journal} {Science}\ }\textbf {\bibinfo {volume}
  {311}},\ \bibinfo {pages} {1293} (\bibinfo {year} {2006})}\BibitemShut
  {NoStop}%
\bibitem [{\citenamefont {Karginov}\ \emph {et~al.}(2010)\citenamefont
  {Karginov}, \citenamefont {Ding}, \citenamefont {Kota}, \citenamefont
  {Dokholyan},\ and\ \citenamefont {Hahn}}]{Karginov2010}%
  \BibitemOpen
  \bibfield  {author} {\bibinfo {author} {\bibfnamefont {A.~V.}\ \bibnamefont
  {Karginov}}, \bibinfo {author} {\bibfnamefont {F.}~\bibnamefont {Ding}},
  \bibinfo {author} {\bibfnamefont {P.}~\bibnamefont {Kota}}, \bibinfo {author}
  {\bibfnamefont {N.~V.}\ \bibnamefont {Dokholyan}}, \ and\ \bibinfo {author}
  {\bibfnamefont {K.~M.}\ \bibnamefont {Hahn}},\ }\href@noop {} {\bibfield
  {journal} {\bibinfo  {journal} {Nat. Biotechnol.}\ }\textbf {\bibinfo
  {volume} {28}},\ \bibinfo {pages} {743} (\bibinfo {year} {2010})}\BibitemShut
  {NoStop}%
\bibitem [{\citenamefont {See Supplemental Material at [{\it URL will be
  inserted by publisher}] for~additional derivations}\ and\ \citenamefont
  {details~of design}()}]{SI}%
  \BibitemOpen
  \bibfield  {author} {\bibinfo {author} {\bibfnamefont {d.}~\bibnamefont {See
  Supplemental Material at [{\it URL will be inserted by publisher}]
  for~additional derivations}}\ and\ \bibinfo {author} {\bibnamefont
  {details~of design}},\ }\href@noop {} {\ }\BibitemShut {NoStop}%
\bibitem [{\citenamefont {Rothemund}(2006)}]{Rothemund06}%
  \BibitemOpen
  \bibfield  {author} {\bibinfo {author} {\bibfnamefont {P.~W.~K.}\
  \bibnamefont {Rothemund}},\ }\href@noop {} {\bibfield  {journal} {\bibinfo
  {journal} {Nature}\ }\textbf {\bibinfo {volume} {440}},\ \bibinfo {pages}
  {297} (\bibinfo {year} {2006})}\BibitemShut {NoStop}%
\bibitem [{\citenamefont {Douglas}\ \emph {et~al.}(2009)\citenamefont
  {Douglas}, \citenamefont {Dietz}, \citenamefont {Liedl}, \citenamefont
  {H{\"o}gberg}, \citenamefont {Graf},\ and\ \citenamefont {Shih}}]{Douglas09}%
  \BibitemOpen
  \bibfield  {author} {\bibinfo {author} {\bibfnamefont {S.~M.}\ \bibnamefont
  {Douglas}}, \bibinfo {author} {\bibfnamefont {H.}~\bibnamefont {Dietz}},
  \bibinfo {author} {\bibfnamefont {T.}~\bibnamefont {Liedl}}, \bibinfo
  {author} {\bibfnamefont {B.}~\bibnamefont {H{\"o}gberg}}, \bibinfo {author}
  {\bibfnamefont {F.}~\bibnamefont {Graf}}, \ and\ \bibinfo {author}
  {\bibfnamefont {W.~M.}\ \bibnamefont {Shih}},\ }\href@noop {} {\bibfield
  {journal} {\bibinfo  {journal} {Nature}\ }\textbf {\bibinfo {volume} {459}},\
  \bibinfo {pages} {414} (\bibinfo {year} {2009})}\BibitemShut {NoStop}%
\bibitem [{\citenamefont {Gillespie}(1977)}]{Gillespie1977}%
  \BibitemOpen
  \bibfield  {author} {\bibinfo {author} {\bibfnamefont {D.~T.}\ \bibnamefont
  {Gillespie}},\ }\href@noop {} {\bibfield  {journal} {\bibinfo  {journal} {J.
  Phys. Chem.}\ }\textbf {\bibinfo {volume} {81}},\ \bibinfo {pages} {2340}
  (\bibinfo {year} {1977})}\BibitemShut {NoStop}%
\bibitem [{\citenamefont {Crooks}(1999)}]{Crooks1999}%
  \BibitemOpen
  \bibfield  {author} {\bibinfo {author} {\bibfnamefont {G.~E.}\ \bibnamefont
  {Crooks}},\ }\href {\doibase 10.1103/PhysRevE.60.2721} {\bibfield  {journal}
  {\bibinfo  {journal} {Phys. Rev. E}\ }\textbf {\bibinfo {volume} {60}},\
  \bibinfo {pages} {2721} (\bibinfo {year} {1999})}\BibitemShut {NoStop}%
\bibitem [{\citenamefont {Seifert}(2011)}]{Seifert2011}%
  \BibitemOpen
  \bibfield  {author} {\bibinfo {author} {\bibfnamefont {U.}~\bibnamefont
  {Seifert}},\ }\href@noop {} {\bibfield  {journal} {\bibinfo  {journal} {Eur.
  Phys. J. E}\ }\textbf {\bibinfo {volume} {34}},\ \bibinfo {pages} {26}
  (\bibinfo {year} {2011})}\BibitemShut {NoStop}%
\bibitem [{\citenamefont {Seifert}(2012)}]{Seifert2012}%
  \BibitemOpen
  \bibfield  {author} {\bibinfo {author} {\bibfnamefont {U.}~\bibnamefont
  {Seifert}},\ }\href@noop {} {\bibfield  {journal} {\bibinfo  {journal} {Rep.
  Prog. Phys.}\ }\textbf {\bibinfo {volume} {75}},\ \bibinfo {pages} {126001}
  (\bibinfo {year} {2012})}\BibitemShut {NoStop}%
\bibitem [{\citenamefont {Machta}(2015)}]{Machta2015}%
  \BibitemOpen
  \bibfield  {author} {\bibinfo {author} {\bibfnamefont {B.~B.}\ \bibnamefont
  {Machta}},\ }\href {\doibase 10.1103/PhysRevLett.115.260603} {\bibfield
  {journal} {\bibinfo  {journal} {Phys. Rev. Lett.}\ }\textbf {\bibinfo
  {volume} {115}},\ \bibinfo {pages} {260603} (\bibinfo {year}
  {2015})}\BibitemShut {NoStop}%
\end{thebibliography}%
\appendix
\section{Construction and setup of the device}
\begin{figure*}
\includegraphics[width=0.7\textwidth]{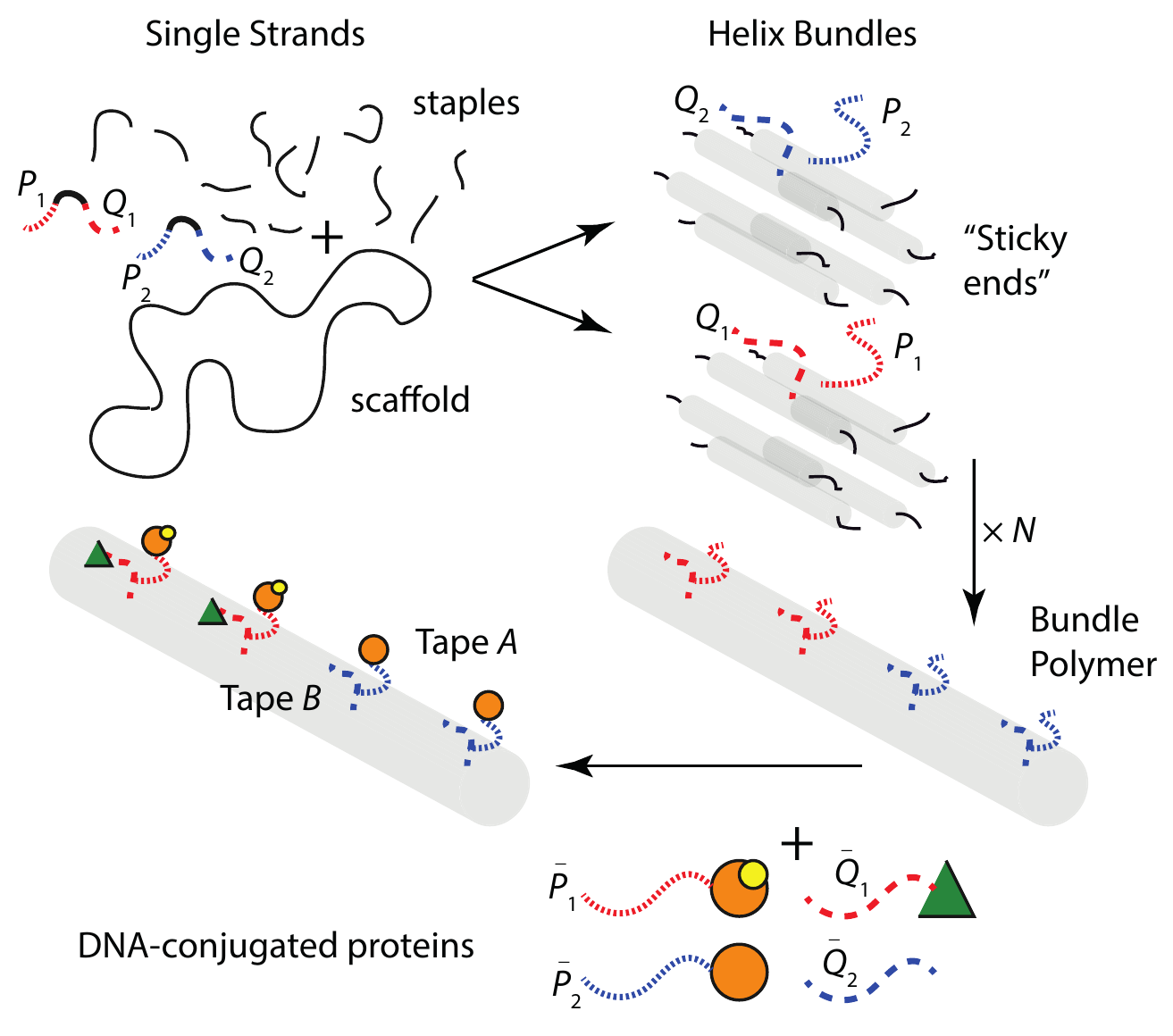}
\caption{Schematic diagram showing the construction of a correlated and coordinated pair of tapes using DNA origami. Firstly, long single-stranded DNA scaffolds are assembled into short helix bundles (the small grey tubes are a single bundle) by short staples with carefully selected sequences \cite{Rothemund06, Douglas09}. Each bundle incoporates one of two staple types that present distinct single-stranded recognition domains.  Bundles then assemble into long polymers (large grey tube) through binding of complementary ``sticky ends". Finally, DNA conjugated proteins are mixed with the polymer to create fully-formed tapes. Biasing the state of the conjugated proteins, as indicated, leads to correlations in the final polymer.  \label{fig:origami}}
\end{figure*}
A proposal for the construction of correlated tapes is illustrated in Fig.\,\ref{fig:origami}.
By mixing single-stranded DNA species with well-designed sequences, a long scaffold strand could be folded into a bundle of helices with ``sticky ends" that are available for subsequent bonding \cite{Rothemund06, Douglas09}. Using the sticky ends, these bundles could then be assembled into long polymers, with the persistence length of these structures tuned through the bundle geometry. Importantly, the polymers consist of repeated units which themselves contain multiple distinct strands, each in a precise location. We propose to manipulate one of these strands in two ways.
\begin{itemize}
\item First, we extend the ends of the strand so that when incorporated into the structure, it possesses two dangling recognition domains $P_1$ and $Q_1$.
\item We introduce a second copy of the strand with distinct recognition domains,  $P_2$ and $Q_2$.
\end{itemize}
When the polymer  is assembled, it will  consist of multiple evenly spaced pairs of recognition domains sites, which will randomly be either $(P_1,Q_1)$ or $(P_2,Q_2)$; the proportion can be tuned by varying the initial strand concentration. Such a configuration is illustrated as the third step in  Fig.\,\ref{fig:origami}.

The substrate protein $X$ could be conjugated with DNA strands via a variety of methods. We propose preparing two separate systems of $X$-conjugated DNA, one with $X$ conjugated to a strand complementary to $P_1$, $\bar{P}_1$, and one with $X$ conjugated to a strand complementary to $P_2$, $\bar{P}_2$. The first system would be exposed to kinases and an ATP dominated bath to drive it to the $X^*$-dominated state, the second system to phosphatases in a phosphate-depleted bath to drive it to the $X$ state. The strands would then be separated from the buffers (perhaps by anchoring them to surface-immobilized origami and washing out the buffer, before breaking the anchorage by displacement). Finally, they are mixed with the wires to generate a random sequence of $X$ and $X^*$ along the wire.

We also propose to conjugate $Y$ with strands of type $\bar{Q}_1$ and $\bar{Q}_2$. In the main text we suggest an activator $Y$ that consists of two molecules, STRAD and MO25  - in this case, the two molecules can be co-localized by conjugation with either end of a strand. We also prepare solutions of $\bar{Q}_1$ and $\bar{Q}_2$ strands with no conjugated $Y$ molecules. The solutions can then be simultaneously mixed with the wires, and $Q$ sites on the wire will be occupied in a manner that depends on the relative frequency of conjugated and un-conjugated $\bar{Q}_1$ and $\bar{Q}_2$, and the excess washed off. 

The result of this process would be a single polymer wire containing two parallel tracks (or tapes), one of which contains $X/X^*$ and the other $Y/\bar{Y}$. If the majority of $\bar{Q}_1$ strands mixed with the wires were conjugated to $Y$, but the majority of $\bar{Q}_2$ strands were not, the presence of $X^*$ on a site will be correlated with the presence of $Y$. Fig.\,\ref{fig:origami} illustrates  this outcome for the case in which this correlation is perfect; alternative choices lead to weaker or reversed correlations. 

The enzyme could be tethered to a surface-immobilised origami, and the tapes  pulled past it in the appropriate bath of nucleotides to operate the device. In the autonomous case, the enzyme and the wire would simply need to be added to solution. Provided the recognition sites that link $Y$ and $X$ to the wire are sufficiently long to allow simultaneous interaction with $E$, but also short enough to prevent cross-talk between sites, the model provided in the text provides an approximate description of device operation. Choosing a relatively stiff polymer wire construction would limit cross-talk.

\section{Solution of the model}
\label{solution}
We now discuss the solution of the model presented in Fig.\,1\,c of the main text. The enzyme is allowed to reach equilibrium with the ATP/ADP bath in between encountering sites $n$ and $n+1$; therefore it is sufficient to analyse each doublet $(A_n, B_n)$ independently. Since, during the interaction window, the state of site $B_n$ implies the activation state of the enzyme, our state space is 12-dimensional (as per Fig. 1c of the main text) and characterised by the  time-dependent probability $p_t(e, a_n, b_n)$, where $e$ is the  state of $E$, $a_n$ is the state of the substrate  $A_n$ and $b_n$ is the state of the activator  $B_n$. Here the possible states of $E$ are $E_-$ (no nucleotide bound), $E_T$ (ATP bound) and $E_D$ (ADP bound), plus activated counterparts, the possible states of $A_n$ are $X$ and $X^*$ and the possible states of $B_n$ are $Y$ and $\bar{Y}$. Working with the vector $p_t(e, a_n, b_n)$ defined as
\begin{equation}
p_t(e, a_n, b_n) = \left( \begin{array}{c} 
p_t(E=E_D, A_n= X, B_n =\bar{Y}) \\
p_t(E=E_-, A_n= X, B_n =\bar{Y}) \\
p_t(E=E_T, A_n= X, B_n =\bar{Y}) \\
p_t(E=E_D, A_n= X^*, B_n =\bar{Y}) \\
p_t(E=E_-, A_n= X^*, B_n =\bar{Y}) \\
p_t(E=E_T, A_n= X^*, B_n =\bar{Y}) \\
p_t(E=E^\dagger_D, A_n= X, B_n ={Y}) \\
p_t(E=E^\dagger_-, A_n= X, B_n ={Y}) \\
p_t(E=E^\dagger_T, A_n= X, B_n ={Y}) \\
p_t(E=E^\dagger_D, A_n= X^*, B_n ={Y}) \\
p_t(E=E^\dagger_-, A_n= X^*, B_n ={Y}) \\
p_t(E=E^\dagger_T, A_n= X^*, B_n ={Y}) 
\end{array}
\right),
\end{equation}
the evolution of the system during the window $t = 0 \rightarrow \tau$ is given by  $p_{t}(e,a_n,b_n) = {\rm e}^{\mathcal{R}t}p_{0}(e,a_n,b_n)$, where  $\mathcal{R}$ is the transition rate matrix
\begin{equation}
\mathcal{R} = \left( \begin{array}{c c}
\mathcal{R}_{\bar Y} & 0 \\
0  & \mathcal{R}_{Y}
\end{array}
\right).
\end{equation}
Here,
\begin{equation}
\mathcal{R}_{\bar Y} = \left( \begin{array}{c c c c c c} 
-1 & 1-\alpha & 0 & 0 & 0 & 0 \\
1 & -2 & 1 & 0 & 0 & 0\\
0 & 1+\alpha & -1 & 0 & 0 & 0  \\
0 & 0 & 0 & -1 & 1-\alpha & 0  \\
0 & 0 & 0 & 1 & -2 & 1 \\
0 & 0 & 0 & 0 & 1+\alpha & -1  \\
\end{array}
\right),
\end{equation}
and 
\begin{equation}
\mathcal{R}_Y = \left( \begin{array}{c c c c c c} 
-1 & 1-\alpha & 0 & 0 & 0 & 0 \\
1 & -2 & 1 & 0 & 0 & 0\\
0 & 1+\alpha & -2 & 1  & 0 & 0  \\
0 & 0 & 1 & -2 & 1-\alpha & 0  \\
0 & 0 & 0 & 1 & -2 & 1 \\
0 & 0 & 0 & 0 & 1+\alpha & -1  \\
\end{array}
\right).
\end{equation}
The initial condition is
\begin{equation}
p_0(e,a_n,b_n) = \\
 \left(\begin{array}{c}
\mathbb{I}_3 (1-\psi)(1-q)\\
\mathbb{I}_3 \psi q \\
\mathbb{I}_3 \psi(1-q) \\
\mathbb{I}_3 (1-\psi)q
\end{array} 
\right)
\left(
\begin{array}{c}
\frac{1-\alpha}{3} \\
1/3\\
\frac{1+\alpha}{3}\\
\end{array}
\right).
\end{equation}
In the period after $\tau$, $A_n$ and $B_n$ are fixed and $E$ relaxes back to its equilibrium distribution according to 
\begin{equation}
p_{t+\tau}(e) = {\rm e}^{\mathcal{R}_E t }p_{\tau}(e),
\end{equation}
with $\mathcal{R}_E$ given by 
\begin{equation}
\mathcal{R}_{E} = \left( \begin{array}{c c c} 
-1 & 1-\alpha & 0  \\
1 & -2 & 1 \\
0 & 1+\alpha & -1  \\
\end{array}
\right),
\end{equation}
and $p_t(e)=(p_t(E=E_D), p_t(E=E_-), p_t(E=E_T))^{\rm T}$. In the main text we rely on the fact that this relaxation reaches steady state to treat each pair of sites $n$ independently, but the details of the dynamics are not that important.

The full behaviour can be solved straightforwardly by identifying the eigenvalues and eigenvectors of $\mathcal{R}$, but the most important results are
\begin{equation}
\begin{array}{c}
p_\tau(A_n=X,B_n=Y) = \vspace{2mm}\\
p_0(B_n=Y) \left( \frac{1-\alpha}{2} +f(\tau)(p_0(A_n=X | B_n=Y) - \frac{1-\alpha}{2}) \right),\\
\\
p_\tau(A_n=X^*,B_n=Y) =  \vspace{2mm}  \\
p_0(B_n=Y) \left( \frac{1+\alpha}{2} +f(\tau)(p_0(A_n=X^* | B_n=Y) - \frac{1+\alpha}{2}) \right),\\
\end{array}
\end{equation}
along with the trivial 
\begin{equation}
\begin{array}{c}
p_\tau(A_n=X,B_n=\bar{Y})=  p_0(A_n=X,B_n=\bar{Y}) = (1-q)(1-\psi),\\
\\
p_\tau(A_n=X^*,B_n=\bar{Y})=  p_0(A_n=X^*,B_n=\bar{Y}) = q\psi.
\end{array}
\end{equation}
 Here, $p_0(B_n=Y) = q(1-\psi) + \psi(1-q)$,  $p_0(A_n=X|B_n=Y) = 1- p_0(A_n=X^*|B_n=Y)   = \psi(1-q)/ (q(1-\psi) + \psi(1-q))$. Finally,
\begin{equation}
f(\tau) = \frac{1}{9} {\rm e}^{-2\tau} \left( 1 + 8 \cosh (\sqrt{3} \tau) + 4 \sqrt{3} \sinh(\sqrt{3}\tau)\right)
\end{equation}
is a function of $\tau$ only that is unity at $\tau=0$ and tends monotonically to zero as $\tau \rightarrow \infty$. The quantities $\Delta I(A_n, B_n)$, $\Delta \tilde{F}(A_n)$, $\Delta \tilde{F}(A_n,B_n)$ and $\langle w_{\rm cycle} \rangle$, as plotted in Fig.\,\ref{fig:work-info-dF-0.5}, all follow directly from the results above.  Specifically, the work done on the chemical bath per cycle is given by the net number of $X^*$ molecules converted into $X$  multiplied by  $\mu_{\rm ATP} - \mu_{\rm ADP}$,
\begin{equation}
\langle w_{\rm cycle}\rangle = kT\ln\left(\frac{1+\alpha}{1-\alpha}\right) \left(q - p_{\tau}(A_{n} = X^*)\right).
\end{equation}
The mutual information $I(A_n;B_n)$ follows directly in terms of the marginal and joint probabilities, and 
\begin{equation}
\tilde{F}(A_n) = - kT H(A_n) = kT \sum_{a_n} p(a_n) \ln p(a_n),
\end{equation}
in this simple case in which $\mu_X = \mu_{X^*}$. Since $B$ does not evolve, $\Delta \tilde{F}(A_n,B_n) =\Delta \tilde{F}(A_n) + k_{\rm B} T I(A_n;B_n)$.

\subsection{Work and Information change for $q=0.5$, $\tau \rightarrow \infty$}
\begin{figure*}
\includegraphics[width=\textwidth]{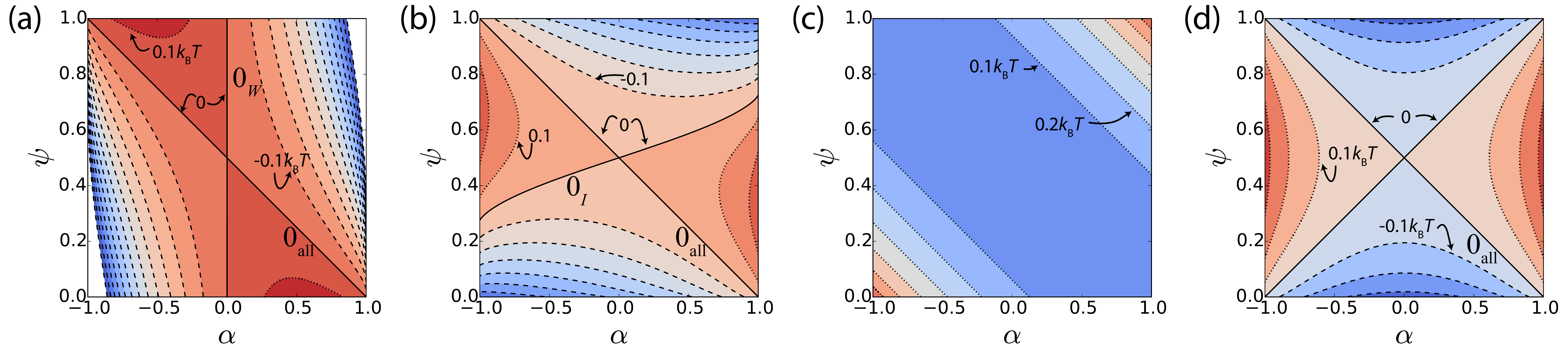}
\caption{Behaviour for $q=0.5$, $\tau \rightarrow \infty$. (a) Work extracted $\langle w_{\rm cycle} \rangle$, with contours running downwards from $0.1 kT$ in units of $0.1 kT$. (c) Information change $\Delta I(A_n;B_n)$, with contours running downwards from 0.2 in units of 0.1.  (c) Free-energy change in tape $A$, $\Delta \tilde{F}(A_n)$, with contours running upwards from $0.1k_{\rm B} T$ in units of $0.1k_{\rm B}T$.  (d) Change in the combined free energy of both tapes  $\Delta \tilde{F}(A_n,B_n)$, with contours running downwards from $0.3k_{\rm B}T$ in units of $0.1k_{\rm B}T$. In all cases, solid contours indicate 0, dotted contours are positive values and dashed contours are negative. \label{fig:work-info-dF-0.5}}
\end{figure*}
In Fig.\,\ref{fig:work-info-dF-0.5} we plot $\langle w_{\rm cycle} \rangle$ and $\Delta I(A_n;B_n)$ for the first case considered in the main text, $q=0.5$, $\tau \rightarrow \infty$. These contour plots imply the regimes indicated in Fig. 2,a of the main text. We also include $\Delta \tilde{F}(A_n)$, which is always positive since $q=0.5$ represents equilibrium between $X$ and $X^*$, $ \tilde{F}(A_n)= F_{\rm eq}(A_n)$. For completeness, we plot $\Delta \tilde{F}(A_n,B_n) = \Delta \tilde{F}(A_n) + k_{\rm B}T \Delta I(A_n;B_n)$.

\subsection{Efficiency}
\begin{figure*}
\includegraphics[width=0.6\textwidth]{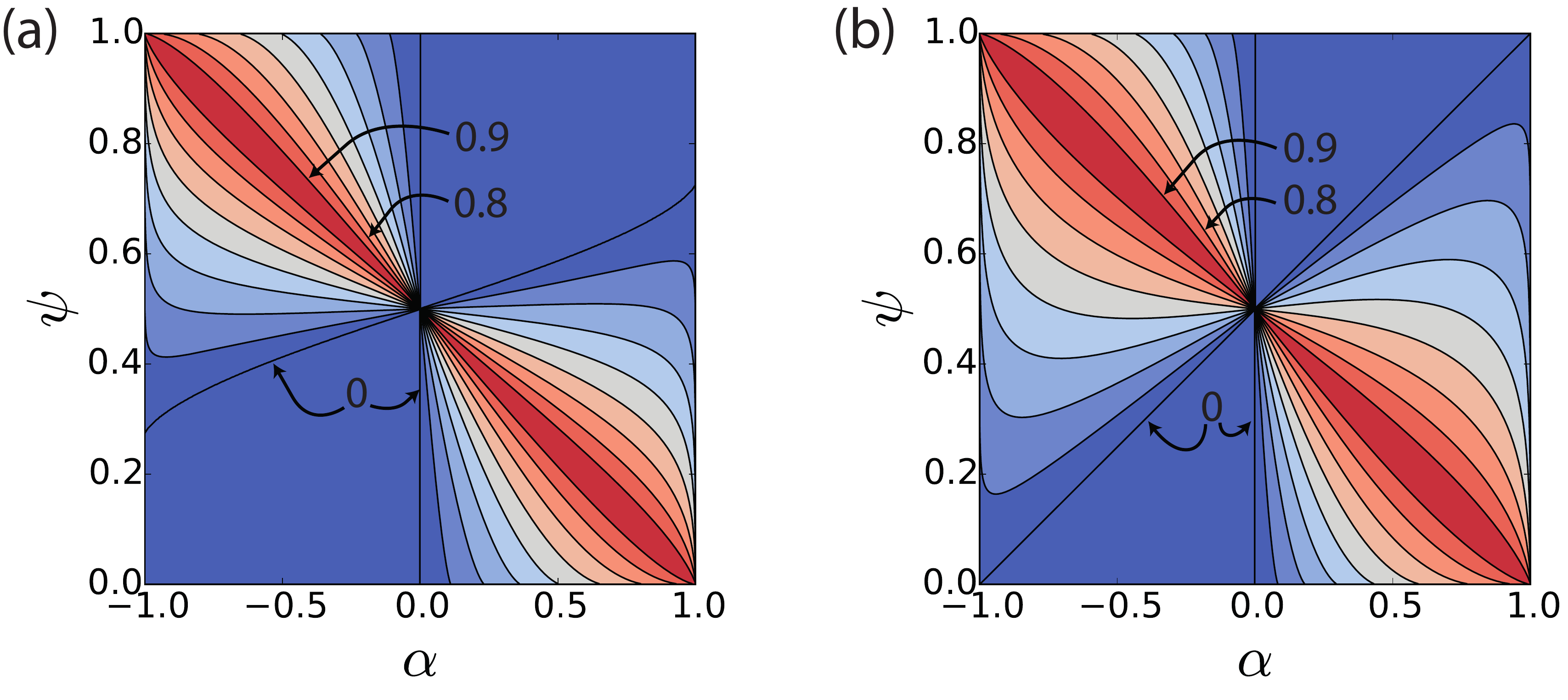}
\caption{Efficiencies (a) $\eta^I_1$ and (b) $\eta^F_1$ for $q=0.5$ and $\tau \rightarrow \infty$. The quantity $\eta^I_1$  represents the  fraction of the information reduction converted into work, or the fraction of chemical work converted into stored information; $\eta^F_1$  is an equivalent quantity in which the entire free energy of the tapes $F(A_n,B_n)$ is taken into account instead of just $I(A_n;B_n)$. In both cases, efficiency is high near the null line $0_b$ where the system is weakly driven and functions close to equilibrium. \label{fig:efficiency_SI}}
\end{figure*}
There are many possible definitions of the efficiency of this device. From the perspective of work and information only, we could consider
\begin{equation}
 \eta^I_1 = 
\begin{cases}
-{<w_{\rm cycle}>}/{kT \Delta I(A_n;B_n)}& {\rm if}\, \langle w_{\rm cycle}\rangle  >0, \\
-{kT \Delta I(A_n;B_n)}/ {<w_{\rm cycle}>}& {\rm if}\, \Delta I({A_n;B_n}) > 0,  \\
0& {\rm otherwise},
\end{cases}
\label{eq:efficiency}
\end{equation}
which is the fraction of the information reduction converted into work, or the fraction of chemical work converted into stored information, respectively. We could also consider the fraction of initially available information that is converted into work,
\begin{equation}
 \eta^I_2 = 
\begin{cases}
{<w_{\rm cycle}>}/{kT I_0(A_n;B_n)}& {\rm if}\,  \langle w_{\rm cycle}\rangle  >0, \\
0& {\rm otherwise}.
\end{cases}
\end{equation}
This approach treats any information remaining between $A_n$ and $B_n$ as waste. The efficiency $\eta^I_2$ is the quantity discussed under the name $\eta$ in the main text. Both  $\eta^I_1$ and $ \eta^I_2$ also treat free energy stored in $A_n$ by the device at the end of the cycle, $\tilde{F}_\tau(A_n)$, as waste. We can define similar efficiencies  $\eta^{F}_1$ and $ \eta^{ F}_2$ in which $\tilde{F}(A_n,B_n)$ replaces $I(A_n;B_n)$ in Eqs. \ref{eq:efficiency}, to treat all free energy within the tapes on an equal footing. 

We plot  $\eta^I_1$and  $\eta^F_1$ for $q=0.5$ in Fig.\,\ref{fig:efficiency_SI} (in this case,  $ \eta^F_2 =  \eta^I_2$, plotted in Fig.\,2\,d of the main text). $\eta^I_1\rightarrow 1$ on the null line $0_b$, when the system deviates only a small amount from equilibrium. However, in this case a vanishingly small amount of the information is used, as illustrated by the plot of  $\eta^I_2$ in the main text. As is evident from Fig.\,\ref{fig:efficiency_SI},  $\eta^F_1$ is similar to but larger than $\eta^I_1$ due to the free energy stored in $A$ not being counted as waste.

\section{Adjusting input parameters}
\subsection{Finite $\tau$}

\begin{figure*}
\includegraphics[width=\textwidth]{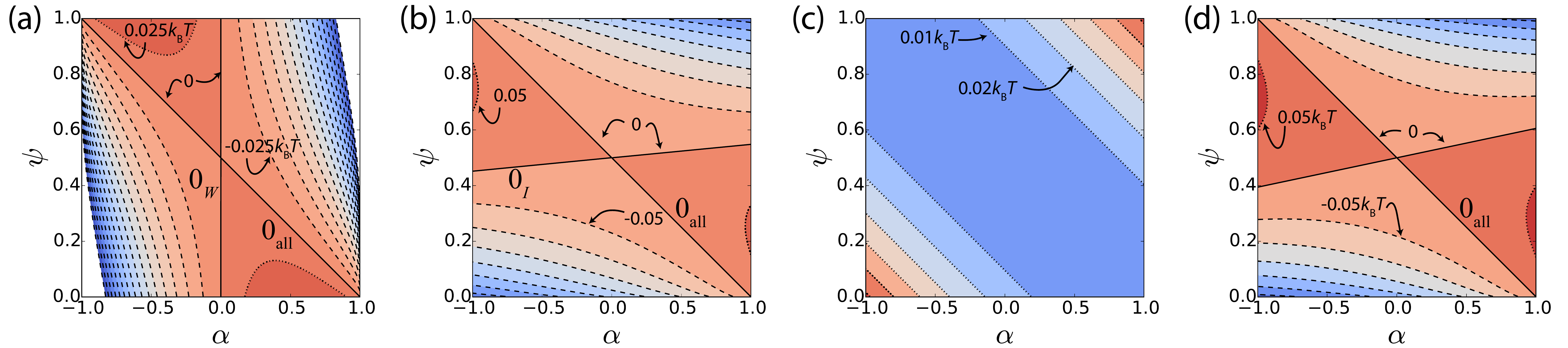}
\caption{Behaviour for $q=0.5$, $\tau =1$, to be compared with Fig.\,\ref{fig:work-info-dF-0.5}. (a) Work extracted $\langle w_{\rm cycle} \rangle$, with contours running downwards from $0.025 kT$ in units of $0.025 kT$. (b) Information change $\Delta I(A_n;B_n)$, with contours running downwards from 0.05 in units of 0.05. (c) Free-energy change in tape $A$, $\Delta \tilde{F}(A_n)$, with contours running upwards from $0.01k_{\rm B} T$ in units of $0.01k_{\rm B}T$.  (d) Change in the combined free energy of both tapes  $\Delta \tilde{F}(A_n,B_n)$, with contours running downwards from $0.05k_{\rm B}T$ in units of $0.05k_{\rm B}T$. In all cases, solid contours indicate 0, dotted contours are positive values and dashed contours are negative.  \label{fig:finite_tau}}
\end{figure*}

We plot $\langle w_{\rm cycle} \rangle$, $\Delta I(A_n;B_n)$, $\Delta \tilde{F}(A_n)$ and $\tilde{F}(A_n, B_n)$ for finite $\tau = 1$ and $q=0.5$ in Fig.\,\ref{fig:finite_tau}. The results are similar to $\tau \rightarrow \infty$ (Fig.\,\ref{fig:work-info-dF-0.5}). Indeed, the lines $0_a$ and $0_b$ are unchanged, as expected. The most notable difference is that, since the reaction during $\tau$ no longer reaches equilibrium, the initial conditions required to lie on the curve $0_c$  change. Thus $I(A_n;B_n)$  is non-monotonic in $\tau$ in some regions of $\alpha-\psi$ space, as illustrated in Fig. 2\,c of the main text. In its region of non-monotonicity, $I(A_n;B_n)$ is initially drained and then restored as the correlations first support and then oppose the thermodynamic drive of the ATP/ADP bath.

\subsection{$q \neq 0.5$}
\begin{figure*}
\includegraphics[width=\textwidth]{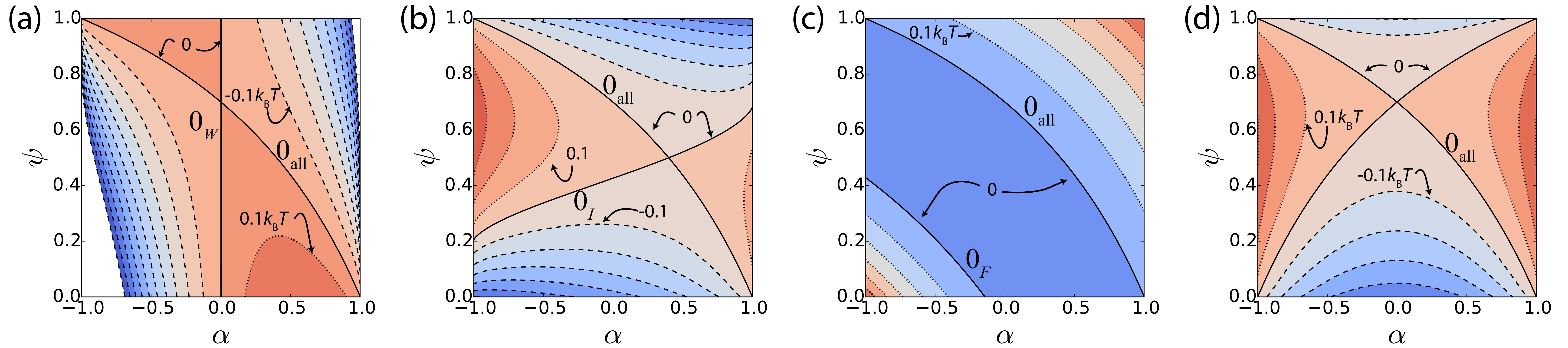}
\caption{Behaviour for $q=0.7$, $\tau \rightarrow \infty$, to be compared with Fig.\,\ref{fig:work-info-dF-0.5}. (a) Work extracted $\langle w_{\rm cycle} \rangle$, with contours running downwards from $0.1 kT$ in units of $0.1 kT$. (b) Information change $\Delta I(A_n;B_n)$, with contours running downwards from 0.3 in units of 0.1. (c) Free-energy change in tape $A$, $\Delta \tilde{F}(A_n)$, with contours running upwards from $0 k_{\rm B}T$ in units of $0.1k_{\rm B}T$.  (d) Change in the combined free energy of both tapes  $\Delta \tilde{F}(A_n,B_n)$, with contours running downwards from $0.2k_{\rm B}T$ in units of $0.1k_{\rm B}T$. In all cases, solid contours indicate 0, dotted contours are positive values and dashed contours are negative. Note the distortions of the null lines compared to   Fig.\,\ref{fig:work-info-dF-0.5}, and the appearance of two contours along which $\Delta \tilde{F}(A_n)=0$, enclosing a region of negative $\Delta \tilde{F}(A_n)$.  \label{fig:work-info-dF-0.7}}
\end{figure*}
In Fig.\,\ref{fig:work-info-dF-0.7}, we plot $\langle w_{\rm cycle}\rangle $ and $\Delta I(A_n; B_n)$ for  $q=0.7$ and $\tau \rightarrow \infty$; the same values used to produce Fig. 3\,a of the main text. We also include $\Delta \tilde{F}(A_n)$ and $\Delta \tilde{F}(A_n,B_n)$. Previously, for $q=0.5$,  the initial tape $A$ was in equilibrium with $\tilde{F}(A_n)= F_{\rm eq}(A_n)$  and so only positive values of 
$\Delta \tilde{F}(A_n)$ were possible (Fig.\,\ref{fig:work-info-dF-0.5}). For $q\neq 0.5$, $\Delta \tilde{F}(A_n)$ is negative between $0_b$ and $0_d$, introducing the new regimes of behaviour illustrated in Fig.\,3\,a of the main text.

The null line $0_b$ is now given by $2\psi(1-q) = (1-\alpha)((1-q)\psi +q(1-\psi))$, and positive work occupies a larger region for $\alpha >0$ (high $[{\rm ATP}]/[{\rm ADP}]$) than $\alpha <0$ since  $q>0.5$ provides an excess of $X^*$ that tends to convert ADP into ATP. The line $0_c$ is shifted; a non-equilibrium tape $A$ can generate $I(A_n;B_n)$ even when a device is fed with no information ($\psi=0.5$) and no imbalance of ATP and ADP ($\alpha=0$).

\subsection{Varying the rates}
\begin{figure*}
\includegraphics[width=\textwidth]{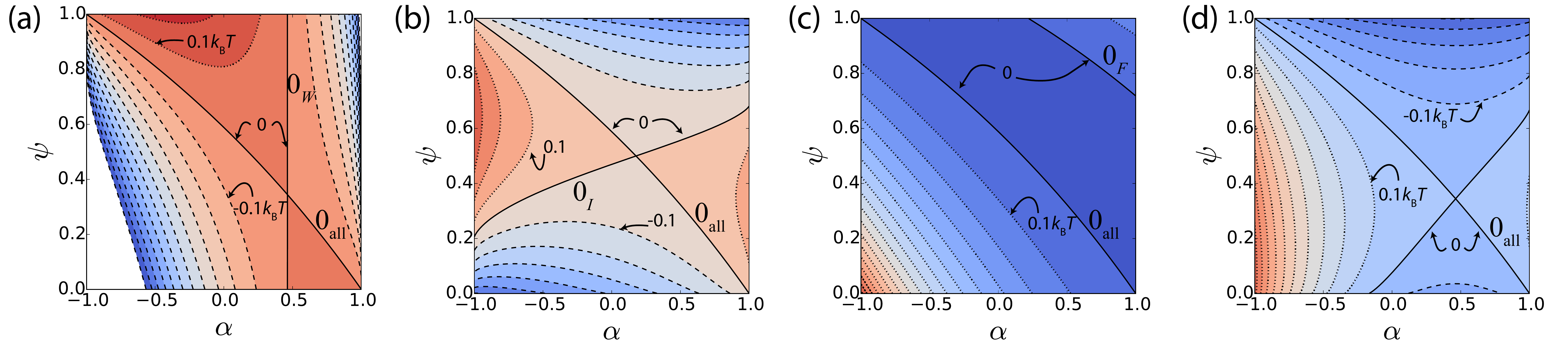}
\caption{Behaviour for $q=0.7$, $\tau \rightarrow \infty$, and varied rates. Specifically,  $k_1^T = k_1^D=1$, $k_{-1}^T=1.4$, $k_{-1}^D=0.8$, $k_2=0.7$ and $k_{-2}=2$ in reduced units. Additionally, we take $\Delta F^0_{\rm D\rightarrow T}=-1$. (a) Work extracted $\langle w_{\rm cycle} \rangle$, with contours running downwards from $0.2 k_{\rm B}T$ in units of $0.1 k_{\rm B}T$. (c) Information change $\Delta I(A_n;B_n)$, with contours running downwards from 0.3 in units of 0.1. (c) Free-energy change in tape $A$, $\Delta \tilde{F}(A_n)$, with contours running upwards from $0 k_{\rm B}T$ in units of $0.1k_{\rm B}T$.  (d) Change in the combined free energy of both tapes  $\Delta \tilde{F}(A_n,B_n)$, with contours running downwards from $1 k_{\rm B}T$ in units of $0.1k_{\rm B}T$. In all cases, solid contours indicate 0, dotted contours are positive values and dashed contours are negative.  Note that although distorted compared to   Fig.\,\ref{fig:work-info-dF-0.7},the overall behaviour is very similar.  \label{fig:varyk}}
\end{figure*}
Instead of setting the majority of rates to be equal, we could consider allowing variability. Retaining the simplification that binding and unbinding of ATP and ADP is  independent of whether the enzyme is active or not gives  the following set of rate constants:
\begin{equation}
\begin{array}{c}
E_-  \underset{k^D_{-1}}{\overset{k^D_1(1-\alpha)}\rightleftharpoons} E_D, \hspace{3mm} E_-  \underset{k^T_{-1}}{\overset{k^T_1(1+\alpha)}\rightleftharpoons} E_T, \\
\\
E^\dagger_D + X^* \underset{k_{-2}}{\overset{k_2}\rightleftharpoons} E^\dagger_T + X.
\end{array}
\end{equation}
Additionally, we set $k^T_{1}  = k^D_1$, which corresponds to assuming the difference between ATP and ADP binding strength is manifest in the off-rate. Finally, although it does not affect the dynamics, it is necessary to identify an intrinsic free-energy change upon interconversion of ATP and ADP (which could, in principle, be set by adjusting the concentration of inorganic phosphate in the bath). The free energy change on converting ATP into ADP is given by $\mu_{\rm ADP} - \mu_{\rm ATP} =  k_{\rm B}T \ln \frac{1-\alpha}{1+\alpha} - \Delta F^0_{\rm D \rightarrow T}$; in the main text $\Delta F^0_{\rm D\rightarrow T}$ was set to zero.  This implies an intrinsic free energy difference between $X^*$ and $X$ of
\begin{equation}
\Delta F^0_{X^* \rightarrow X} = \mu_X - \mu_{X^*} =  - \Delta F^0_{\rm D \rightarrow T} - k_{\rm B}T \ln \left( \frac{k_2 k^T_{-1}}{k_{-2} k^D_{-1}} \right),
\end{equation} 
which was zero under the assumptions of the main text. As a result, unlike the main text and Section \ref{solution}, the free energy of tape $A$, $\tilde{F}(A_n)$, is no longer specified purely by its entropy, but contains a contribution from the relative stabilities of $X$ and $X^*$:
\begin{equation}
\begin{array}{c}
\tilde{F}(A_n) - \tilde{F}_{\rm eq}(A_n) = \vspace{1mm}\\ \left( p(A_n=X^*) - p_{\rm eq}(A_n = X^*) \right) (\mu_{X^* }- \mu_{X})  - k_{\rm B}T H(A_n).
\end{array}
\end{equation}
Similarly, the work done on the chemical bath contains a term due to the intrinsic stability of ATP and ADP
\begin{equation}
\langle w_{\rm cycle} \rangle = \left(q - p_\tau(A_n=X^*)\right) \left( \Delta F^0_{\rm D \rightarrow T} + k_{\rm B}T \ln \left( \frac{1+\alpha}{1-\alpha} \right) \right).
\end{equation}
The time evolution of the system can be solved exactly as in Section \ref{solution}, and equivalent quantities are required to calculate $\Delta I(A_n; B_n)$, $\langle w_{\rm cycle} \rangle $, $\Delta \tilde{F}(A_n)$ and $\Delta \tilde{F}(A_n,B_n)$. We plot these quantities as a function of $\alpha$ and $\psi$ for $q=0.7$ in Fig.\,\ref{fig:varyk}, with the rate constants specified in the caption. Although the shape is distorted, the regimes and dividing lines identified in our earlier analysis still appear. Indeed, the only qualitative difference between Fig.\,\ref{fig:work-info-dF-0.7} and Fig.\,\ref{fig:varyk} is that the contour $0_d$ (a line along which $\Delta \tilde{F}(A_n)=0$), now appears on the top right rather than the bottom left of the figure. In the new setting, inverting the probabilities of $X$ and $X^*$ does not preserve $\tilde{F}(A_n)$, since $X$ and $X^*$ have different intrinsic stabilities. In fact, for the parameters we have chosen, $X^*$ is more stable than $X$. Therefore the line $0_d$, with  $\Delta \tilde{F}(A_n)=0$, is found at an even larger excess of $X^*$ than initially, with the decrease in $H(A)$ compensated by the increased intrinsic stability of molecules attached to tape $A$.

 Importantly, even with unequal rates, the  existence of mutual information between $A$ and $B$ always implies a thermodynamic resource $k_BTI(A_n; B_n)$ that can be exploited.  It is also always possible to chose a value of $q$ such that  $\tilde{F}(A_n) = F_{\rm eq}(A_n)$, and the information is the only resource of the tapes. For our parameters in Fig.\,\ref{fig:varyk}, this value is $q \approx 0.82$.  By contrast, the ``information entropy" $H(A_n)$ that is exploited in many other hypothetical machines \cite{Mandal2012,Mandal2013,Cao2015} has a less important role when rate constants are varied. In particular, $H(A_n) < H_{\rm max}(A_n)$ does not necessarily imply a reservoir from which work can be extracted, and tape $A$ can either increase of decrease $H(A)$ as it does work on its environment.

\section{Explicit verification of fluctuation relations during $\tau$}
 Trajectories $z(t)=(e(t),a_n(t),b_n(t))$ within the window $0 \rightarrow \tau$ generate a stochastic entropy $\sigma[z(t),p_t(z)]$ that has two contributions: one from the details of trajectory itself and another from the overall evolution of the probability distribution of the system, $p_t(z)$ \cite{Crooks1999,Seifert2012}. 
\begin{equation}\label{eq:tot_s}
\sigma[z(t),p_t(z)]= k_{\rm B} \ln\frac{p[z(t)|z_0]}{p^*[z^*(t)|z_{\tau}]}  - k_{\rm B}\ln\frac{p_{\tau}(z_{\tau})}{p_{0}(z_{0})},
\end{equation}
where $p^*[z^*(t)|z_{\tau}]$ is the probability of the time-reversed trajectory occuring under time-reversed protocols, given a starting point equal to the end-point of the forward trajectory $z_\tau$.
The first term can be further subdivided into \cite{Seifert2012, Seifert2011}
\begin{equation}
k_{\rm B} \ln\frac{p[z(t)|z_0]}{p[z^*(t)|z_{\tau}]} = -\frac{Q_{\rm in}}{T} + \Delta s_{0, \tau}^{\rm int},
\end{equation}
where $Q_{\rm in}$ is the total heat input to the combined enzyme, tape and bath system, and  $ \Delta s_{0, \tau}^{\rm int}$ is the increase in  intrinsic entropy of the states of the enzyme, tapes and bath between the initial and final states. The intrinsic entropy is a property of the chemical macrostate, and arises from coarse graining \cite{Seifert2011}. 

The absence of a time-dependent control means that $Q_{\rm in}$  is given solely by the change in internal energy of the combined system of enzymes, tape and bath, $\Delta \mathcal{E}_{0,\tau}$.  Thus in this special case, the entropy generated by $z(t)$ is specified purely by the initial and final states of the system
\begin{equation}\label{eq:tot_s2}
\sigma[z(t),p_t(z)]= -\frac{\Delta \mathcal{E}_{0,\tau}}{T} + \Delta s^{\rm int}_{0,\tau} - k_{\rm B}\ln\frac{p_{\tau}(z_{\tau})}{p_{0}(z_{0})},
\end{equation}
As a consequence, the fluctuation relation $\langle \exp(-\sigma[z(t),p_t(z)]/k_{\rm B}) \rangle =1$ can be verified simply by summing $\exp(-\sigma[z(t),p_t(z)]/k_{\rm B})$ over all possible combinations of initial and final states, weighted by the probabilities of those combinations occurring (which follow from the solution of the system in Section \ref{solution}). As noted in the main text, in our case this fluctuation relation is equivalent to $\langle {\rm e}^{-\bar{\sigma}[e_n(t),a_n(t),{p}_t(e,a_n)]/k_{\rm B} + \Delta i_{0,\tau}(b_n;e,a_n)}\rangle = 1$ \cite{Sagawa2012}, where $i(b_n;e,a_n) = \ln(p(b_n,e,a_n)/p(b_n)p(e,a_n))$ is the pointwise mutual information between $B_n$ and $A_n, E$ and 
\begin{equation}
\bar{\sigma} [e_n(t),a_n(t),{p}_t(e,a_n)]=  -\frac{\Delta \mathcal{E}_{0,\tau}}{T} + \Delta s^{\rm int}_{0,\tau} - k_{\rm B} \ln \frac{p_{\tau}(e_{\tau}, {a_n}_\tau)}{p_{0}(e_0, {a_n}_0)} 
\end{equation}
is the entropy change due to reactions involving $E$, $A_n$ and the chemical bath. This relation holds since
\begin{equation}
k\ln\frac{p_{\tau}(z_{\tau})}{p_{0}(z_{0})} = k_{\rm B} \ln \frac{p_{\tau}(e_{\tau}, {a_n}_\tau)}{p_{0}(e_0, {a_n}_0)}  +k_{\rm B}  \Delta i_{0,\tau}(b_n;e,a_n),
\end{equation}
noting that $B_n$ does not change during the dynamics

There are 12 possible initial states and 12 possible final states to sum over. The dynamics does not connect initial and final states with different $B_n$, which leaves 72 distinct combinations of initial and final states. Further, states of different $A_n$ and $B_n= \bar{Y}$ are unconnected, leaving only 54 combinations, which can be trivially summed over using {\it eg.} Mathematica to demonstrate that $\langle \exp(-\sigma[z(t),p_t(z)]/k_{\rm B}) \rangle =\langle {\rm e}^{-\bar{\sigma}[e_n(t),a_n(t),{p}_t(e,a_n)]/k_{\rm B} + \Delta i_{0,\tau}(b_n;e,a_n)}\rangle = 1$.

For the case of infinite $\tau$, and assuming that manipulation of the tapes is performed quasistatically, this is the only entropy generation from the system during operation. For finite $\tau$, the enzyme relaxes after the interaction window, the fluctuation theorem for which can also be verified explicitly. 

\section{Subtleties relating to external manipulation of the tapes}
The operation of the device requires the tapes to be slowly moved past the enzyme. This could be done continuously, or in stages (move, stop, move again). In either case the modelling of the interactions as constant for a fixed time $\tau$ is only approximate, although it is unlikely that lifting this approximation would provide fundamentally different  physics.

If the tapes are pulled at a finite rate, work will be done in the attachment/detachment of proteins as they move past the enzyme that is not included in our analysis. This is particularly true for the $Y-E$ interaction, which is assumed to be reasonably strong. If this dissipation is to be minimised through slow manipulation of the tapes, it sets requirements on the  intrinsic stability of the phosphorylation states of $X$ as well as the DNA structure (as does the construction process). We note that these requirements are only evident when the interaction between information and the device is made physical, as in this case but not in prior work .

As has been argued elsewhere, infinitely slow but unidirectional motion implies finite dissipation when the system that causes unidirectional motion is taken into account \cite{Machta2015}. If multiple tapes are manipulated simultaneously, however, this required dissipation does not scale extensively with the system size. Although in principle forces can be applied to biochemical systems using, for example, electric fields, in practice manipulation is often achieved using devices such as optical tweezers. In this case the manipulating device  is enormously dissipative, a factor that is usually neglected. The possibility of eliminating this dissipation, along with the need to manipulate tapes slowly and the consequent requirements for stability of the system, make autonomous devices attractive.

\section{Subtleties related to averaging the mutual information over trajectories}
In the main text we introduced an autonomous device that could interact with a number of $(A_n,B_n)$ pairs without being manipulated. In Fig.\,3\,c we plotted the time evolution of the mutual information between pairs of sites, obtained by averaging $p_t(a_n,b_n)$, $p_t(a_n)$ and $p_t(b_n)$ over multiple simulations and then calculating $I(A_n;B_n)$. Also shown are the values of $I(A_n;B_n)$ obtained within individual stochastic simulations, calculated by estimating $p_t(a_n,b_n)$,  $p_t(a_n)$ and $p_t(b_n)$ by sampling from many equivalent pairs of sites within state $z(t)$ of the simulation, and identifying {\it e.g.} $p_t(a_n)$ as the fraction of $A$ in state $X^*$ at time $t$.

Formally, although the estimates in the second case have means equal to $p_t(a_n,b_n)$,  $p_t(a_n)$ and $p_t(b_n)$  respectively, their fluctuations about the means are correlated. Thus averaging the apparent mutual information -- a non-linear function of probabilities -- in each individual simulation does not give the true mutual information  between $A_n$ and $B_n$. $I(A_n;B_n)$ is the physically meaningful quantity from which work can be extracted, since the average of the trajectory-dependent information associates information with random fluctuations in equilibrium, which of course could only be exploited through additional measurement and correlation.

This subtlety does not arise for the work, for which the physically meaningful quantity is the average of the work obtained in each individual simulation. In practice, however, we observe that the differences between the average of the trajectory-dependent information and  the true $I(A_n;B_n)$ are relatively small unless  $I(A_n;B_n) \approx 0$ or the total number of pairs is very low.

\end{document}